\documentclass{aa}  

\usepackage{graphicx}
\usepackage{array}
\usepackage{tabularx}
\usepackage{txfonts}
\usepackage{hyperref}
\def\ltsima{$\; \buildrel < \over \sim \;$}
\def\simlt{\lower.5ex\hbox{\ltsima}}
\def\gtsima{$\; \buildrel > \over \sim \;$}
\def\simgt{\lower.5ex\hbox{\gtsima}}

\def\msun{{\rm M_{\odot}}}

\def\be{\begin{equation}}
\def\ee{\end{equation}}

\def\del#1{{}}
\def\ltsima{$\; \buildrel < \over \sim \;$}
\def\simlt{\lower.5ex\hbox{\ltsima}}
\def\gtsima{$\; \buildrel > \over \sim \;$}
\def\simgt{\lower.5ex\hbox{\gtsima}}

\usepackage{xcolor}

\newcolumntype{L}{>{$}l<{$}}
\newcolumntype{C}{>{$}c<{$}}
\newcolumntype{R}{>{$}r<{$}}

\usepackage{booktabs}
\usepackage{siunitx}
\usepackage{comment}

\begin{document}

   \title{Simple relations from complex outflows: How the $M-\sigma$ relation emerges in a multi-phase environment}
   \titlerunning{$M-\sigma$ in multi-phase outflows}

   \author{M. Tartėnas\orcid{0009-0006-7373-180X}\inst{1}, K. Zubovas\orcid{0000-0002-9656-6281}\inst{1,2} \and E. Skuodas\orcid{0000-0000-0000-0000}\inst{2}}

   \institute{Center for Physical Sciences and Technology, Saulėtekio al. 3, Vilnius LT-10257, Lithuania\\
              \email{kastytis.zubovas@ftmc.lt}
            \and
             Astronomical Observatory, Vilnius University, Saulėtekio al. 3, Vilnius LT-10257, Lithuania\\ 
             }

   \date{Received ...; accepted ...}

  \abstract
   {
   The tight empirical $M-\sigma$ relation between the mass of a supermassive black hole (SMBH) and the velocity dispersion of the host galaxy bulge is often interpreted as the result of self-regulation via active galactic nucleus (AGN) feedback. This picture is motivated by analytical and semi-analytical models in which momentum-driven AGN winds can expel the gas once the SMBH reaches a critical mass. However, these models typically assume idealised conditions: smooth gas distributions, spherical symmetry, and very efficient cooling of the shocked AGN wind. It is unclear whether this paradigm is applicable under more realistic conditions.
   }
   {
   We checked whether AGN outflows can establish the $M-\sigma$ relation in a multi-phase and turbulent galactic bulge subject to realistic radiative cooling while conserving the shocked AGN wind energy.
   }
   {
   We calculated ran a suite of purpose-built hydrodynamical simulations of AGN outflows in turbulent gas shells, covering a wide range of constant AGN luminosities. We tracked the outflow evolution over the course of $\geq1$~Myr. We analysed the effect of AGN outflow on the cold dense gas and SMBH feeding, estimating the luminosity threshold for removing most of the cold gas from the central regions.
   }
   {
   We find that AGNs with significantly sub-Eddington luminosities cannot suppress SMBH feeding, while luminosities exceeding $\sim 0.7$ times Eddington clear out both the diffuse hot gas and the cold clumps, consistent with the momentum-driven outflow formalism. We also show that dense gas clusters are affected almost exclusively by the AGN wind momentum, while the shocked wind energy escapes through low-density channels and inflates large bubbles of diffuse gas.
   }
   {
   Active galactic nucleus wind-driven energy-conserving feedback in a turbulent multi-phase medium affects the dense gas only via the wind momentum. Thus, the momentum-driven outflow paradigm is applicable for explaining the $M-\sigma$ relation even in realistic systems.
   }

   \keywords{black hole physics -- 
            ISM: general, jets and outflows --
            galaxies: active, general --
            (galaxies:)quasars: general 
               }

   \maketitle
%

\section{Introduction}

It has been known for more than two decades that the masses of supermassive black holes (SMBHs) correlate with the properties of their host galaxies \citep[e.g.][]{Magorrian1998AJ, Ferrarese2000ApJ}. Of the numerous relations proposed so far, the $M-\sigma$ relation, which links SMBH mass to the velocity dispersion of the host-galaxy spheroid \citep{Ferrarese2000ApJ, Gultekin2009ApJ, McConnell2013ApJ, Saglia2016ApJ, deNicola2019MNRAS}, is the most fundamental \citep{Shankar2013MNRAS, Marsden2020FrP}. Its existence is readily explained as the result of active galactic nucleus (AGN) feedback. The energy released via luminous accretion onto the SMBH during the AGN phase couples to the interstellar medium (ISM) and regulates the growth of both the SMBH and the host galaxy, establishing various correlations between the SMBH mass and galaxy properties. Feedback manifests in the form of radiation, narrow jets, and/or wide-angle outflows \citep{Fabian2012ARAA, King2015ARAA, Harrison2024Galax}. Of these, the wide-angle wind-driven outflow model (see\citealt{King2015ARAA, Zubovas2019GReGr} for recent reviews) is the most successful in explaining galaxy-scale effects, such as the properties of observed outflows, their influence on galaxy properties, and the $M-\sigma$ relation.

Within the framework of AGN wind-driven feedback, AGN luminosity ($L_{\rm AGN}$) is communicated to the surrounding gas via a quasi-relativistic wind emanating from the accretion disc \citep{King2010MNRASa}. The wind moves with velocities ($v_{\rm w}$) of $\sim 0.1c$ and carries a kinetic power ($\dot{E}_{\rm w})$ of $\sim 0.05 L_{\rm AGN}$. Upon encountering the relatively static ISM, the wind develops a strong shock with a post-shock temperature ($T_{\rm sh})$ of $\sim 10^{10}$~K. The pressure in the shocked wind bubble is significantly higher than that of the ISM, so the bubble begins to expand. At the same time, the inverse-Compton process --- the only efficient cooling process at this temperature --- cools the shocked gas. Depending on which of the two processes --- expansion or cooling --- operates on a shorter timescale, the resulting outflow can be either momentum- or energy-driven. An energy-driven outflow is approximately adiabatic and carries most of the wind energy ($\dot{E}_{\rm kin, out}$), i.e. $\sim 0.02 L_{\rm AGN}$ (the rest of the wind energy is used up to do work against gravity and $p$d$V$ work). This results in a large-scale ($R > 1$~kpc), fast ($v_{\rm out} \sim 1000$~km~s$^{-1}$), and massive ($\dot{M}_{\rm out} \sim 1000 \, \msun$~yr$^{-1}$) outflow. These properties agree well with those of observed outflows in AGN host galaxies \citep{Zubovas2012ApJ, Cicone2014AA, Fluetsch2019MNRAS, Lutz2020AA}. The momentum-driven outflow carries only the direct momentum of the wind, $\dot{p}_{\rm out} \sim L_{\rm AGN}/c,$ and is almost two orders of magnitude less powerful. The condition for this type of outflow to overcome the gravitational potential of the host galaxy leads to the $M-\sigma$ relation \citep{King2010MNRASa}.

How do we reconcile the fact that the $M-\sigma$ relation requires transferring only the wind momentum to the gas, while large-scale outflows appear to be driven by adiabatic expansion of the same shocked wind bubble? At least two solutions exist. The first suggests that the wind cools efficiently close to the SMBH, where the AGN radiation field is stronger \citep{King2003ApJ, King2010MNRASa}, and transitions to an adiabatic state beyond the cooling radius ($R_{\rm cool}$) of $\sim 0.5$~kpc. Whether this happens is a matter of some debate: although at least one galaxy, NGC 4051, shows a spectral feature that resembles the expected signature of the cooling wind \citep{Pounds2011MNRAS}, in general it is not found in AGNs \citep{Bourne2013MNRAS}.

The other possibility is that the outflow type depends on the density of impacted gas. If the ions and electrons in the shocked wind form a two-temperature plasma, the cooling rate diminishes significantly and the cooling radius reduces to $<1$~pc \citep{Faucher2012MNRASb}. This is consistent with most observed AGN spectra \citep{Bourne2013MNRAS}. However, the outflow expands primarily through low-density channels in the clumpy ISM, leaving cool dense gas behind \citep{Zubovas2014MNRASb, Ward2024MNRAS}. The dense gas is then affected primarily by the wind momentum, and the condition that this gas must be pushed away establishes the $M-\sigma$ relation.

This description is qualitative because quantitative analytical calculations are impossible when one considers a multi-phase, turbulent, and non-spherical ISM. In a recent paper \citep[hereafter Paper I]{Zubovas2024AA}, we investigated the effects of turbulence and cooling on energy-driven outflows and found that the inclusion of cooling leads to the formation of multi-phase outflows, where the dense gas may be affected mostly by the wind momentum rather than its energy. Here, we extend the analysis of dense gas motion in outflows driven by AGNs with different luminosities. We show that AGNs with luminosities well below the Eddington limit are unable to prevent the significant accretion of cold dense gas, while those with $L_{\rm AGN} > L_{\rm Edd}$ efficiently remove cold gas and stifle further SMBH growth, as predicted by the momentum-driven outflow formalism. We also show that dense gas clusters mostly experience feedback consistent with pure momentum driving.

We briefly review the analytical derivation of outflow parameters in Sect. \ref{sec:analytical}, with an emphasis on the $M-\sigma$ relation and conditions for the removal of cold gas. We present the numerical setup in Sect. \ref{sec:sims} and the simulation results in Sect. \ref{sec:results}. We discuss the implications of our results in Sect. \ref{sec:discuss}, focusing on establishing and maintaining the $M-\sigma$ relation across cosmic time, the expected properties of multi-phase outflows, and the importance of gas self-gravity and magnetic fields. We summarise the main results and conclude in Sect. \ref{sec:sum}.

\section{The $M-\sigma$ relation in different kinds of outflows}
\label{sec:analytical}

In the momentum-conserving regime, the equation of motion of a spherical outflow shell at distance $r$ is \citep[following][]{King2010MNRASa,King2015ARAA}
\begin{equation}\label{eq:eom_momentum}
\begin{split} 
    \frac{{\rm d}p_{\rm out}}{{\rm d}t} &\equiv \frac{{\rm d}}{{\rm d}t}\left[M_{\rm g}\left(r\right)\dot{r}\right] = \\&= \frac{L_{\rm AGN}}{c} - \frac{G M_{\rm g}\left(r\right)\left[M_{\rm BH}+M_{\rm b}\left(r\right) + M_{\rm g}\left(r\right)/2\right]}{r^2},
\end{split}
\end{equation}
where $M_{\rm g}(r)$ is the mass of the swept-up gas within the outflow radius $r$, $M_{\rm BH}$ is the SMBH mass and $M_{\rm b}$ and $M_{\rm g}$ represent the non-gaseous and gaseous components of the distributed mass of the host, all measured within the outflow radius. Assuming that the AGN luminosity is equal to the Eddington luminosity $L_{\rm Edd} = 4\pi G M_{\rm BH} c/\kappa$, where $\kappa \approx 0.4$~cm$^2$~g$^{-1}$ is the electron scattering opacity, and that the gas and total matter are distributed isothermally, with $M\left(r\right) = 2 \sigma^2 r/G$, $M_{\rm g}\left(r\right) = f_{\rm g}M\left(r\right)$ and $M_{\rm b}\left(r\right) = \left(1-f_{\rm g}\right)M\left(r\right)$, we can show that the outflow can only expand to an arbitrarily large radius if the SMBH mass exceeds
\begin{equation} \label{eq:msigma}
    M_\sigma = \frac{f_{\rm c}\kappa}{\pi G^2} \sigma^4 \approx 3.8\times10^8 \sigma_{200}^4 \, \msun;
\end{equation}
here, $f_{\rm c} = 0.16$ is the cosmological baryon fraction and we scale the velocity dispersion to $\sigma_{\rm 200} \equiv \sigma/(200\, {\rm km \, s}^{-1})$. At large radii, the outflow velocity can be expressed as
\begin{equation} \label{eq:dotr_mom}
    \dot{r}^2 \approx 2\sigma^2\left(\frac{lm}{f}-1\right),
\end{equation}
where $m \equiv M_{\rm BH}/M_\sigma$,  $f \equiv f_{\rm g}/f_{\rm c}$ is the current gas fraction scaled to the cosmological value, and $l \equiv L_{\rm AGN}/L_{\rm Edd}$. This equation clearly has no solution unless $lm/f > 1$, i.e. the outflow can only expand to large radii if the black hole mass is large enough, the AGN is super-Eddington or the gas density is low.

The kinetic power of the outflow is
\begin{equation}
    \dot{E}_{\rm kin,m} = \frac{\dot{M}_{\rm out}\dot{r}^2}{2} \lesssim \frac{L_{\rm AGN}}{2} \frac{\dot{r}}{c},
\end{equation}
where we used the upper limit $\dot{M}_{\rm out}\dot{r} \lesssim L_{\rm AGN}/c$ obtained from Eq. \ref{eq:eom_momentum} by neglecting gravity. Using Eq. (\ref{eq:dotr_mom}), the fraction of AGN luminosity transferred to the outflow is then
\begin{equation}
    \frac{\dot{E}_{\rm kin,m}}{L_{\rm AGN}} \approx \frac{\sigma}{\sqrt{2}c} \sqrt{\frac{lm}{f}-1} \approx 5\times10^{-4} \,\sigma_{\rm 200} \sqrt{\frac{lm}{f}-1}.
\end{equation}
In the energy-driven regime, the equation of motion is similar, but the driving term is replaced with $4\pi r^2 P$, where $P$ is the pressure of the shocked wind bubble. It is calculated using the energy equation
\begin{equation}
    \frac{{\rm d}}{{\rm d}t}\left(\frac{PV}{\gamma-1}\right) = \frac{\eta}{2}L_{\rm AGN} - P\frac{{\rm d}V}{{\rm d}t}- \frac{GM_{\rm g}\left(M_{\rm b}+M_{\rm g}/2\right)}{r^2}\dot{r},
\end{equation}
where $\gamma$ is the specific heat ratio of the gas. We have dropped the $(r)$ notation for brevity. The derivation of the full equation of motion is algebraically involved, but straightforward; we refer the reader to Appendix A of \citet{Zubovas2022MNRAS} for details. In its general form, the equation of motion has no analytical solutions. If we make the same simplifying assumptions regarding the gas distribution as before, a constant-velocity solution emerges \citep[cf.]{King2015ARAA}:
\begin{equation} \label{eq:dotr_en}
    v_{\rm e} \equiv \dot{r} \approx \left(\frac{2 \eta l \sigma^2 c}{3} \frac{f_{\rm c}}{f_{\rm g}}\frac{M_{\rm BH}}{M_\sigma}\right)^{1/3} \approx 925 (lm/f)^{1/3} \sigma_{200}^{2/3} \, {\rm km\, \rm s}^{-1}.
\end{equation}
The kinetic power of the outflow is $\sim 1/3$ of the total energy injected into the ISM, i.e.
\begin{equation}
    \frac{\dot{E}_{\rm kin, e}}{L_{\rm AGN}} \approx 0.016,
\end{equation}
which is $\sim 30/\sigma_{200}$ times higher than in the momentum-driven case. It is clear that an outflow driven by all of the AGN wind energy can escape from the galaxy much more easily, and so the SMBH growth would be quenched at a much lower mass. This unrealistic outcome can be prevented in several ways. The AGN can always be significantly sub-Eddington, although this appears unlikely \citep{King2010MNRASb, Mountrichas2024A&A}. The gas can be very dense, with $f \gg 1$, but this leads to rapid fragmentation into stars. Finally, the ISM can be `porous' \citep{Silk1998AA}, allowing the energy of the outflow to leak out through low-density channels \citep{Zubovas2014MNRASb, Ward2024MNRAS}. Cold gas clumps left behind the fast outflow then experience both a confining pressure of being embedded in a shocked wind bubble and a combined drag and directional pressure force from the wind. These forces are of order $\rho_{\rm w} v_{\rm w}^2$ per unit cross-sectional area of the cold gas clump \citep{King2015ARAA}, which is precisely the momentum rate of the AGN wind. So it is possible that dense gas is affected primarily, or even solely, by the AGN wind momentum, while the energy of the shocked wind is carried by the outflow of diffuse gas.

\section{Numerical simulations} \label{sec:sims}

Our numerical setup is very similar to that of turbulent cooling simulations of Paper I. To recap, we used a modified version of GADGET-3 \citep{Springel2005MNRAS} with the SPHS formulation of the main hydrodynamical equations \citep{Read2010MNRAS, Read2012MNRAS} and a Wendland kernel \citep{Wendland95, Dehnen2012MNRAS} with particle neighbour number $N_{\rm ngb} = 100$. The initial conditions consist of an SMBH and a turbulent gas shell embedded in a static background gravitational potential. The SMBH particle has a mass of $M_{\rm BH} = 10^8 \, \msun$ and is fixed at the centre of the simulation. Using Eq. \ref{eq:msigma}, we derived the velocity dispersion $\sigma_{\rm b} = 142$~km~s$^{-1}$ and used it to generate the background potential. The potential corresponds to an enclosed mass $M_{\rm b}\left(<1\,{\rm kpc}\right) = 9.4\times10^9 \, \msun$ but extends indefinitely far. The gas is initially distributed in a shell between $r_{\rm in} = 0.1$~kpc and $r_{\rm out} = 1$~kpc. The total mass is $M_{\rm g}\left(<1 \, {\rm kpc}\right) = 9.4\times 10^{8} \msun = 0.1 M_{\rm b}\left(<1 \, {\rm kpc}\right)$, i.e. our modelled system has a gas fraction $f_{\rm g} = 0.1$. A real gas-rich bulge in a galaxy with a $10^8 \, \msun$ SMBH would extend farther and be more massive, but we are only interested in gas dynamics within the central several hundred parsecs, so truncating the gas distribution closer in does not affect our results. The gas is tracked with $N \approx 1.08\times 10^{6}$ smoothed-particle hydrodynamics (SPH) particles (the exact number differs in simulations with different initial conditions, as explained below), giving a particle mass $m_{\rm SPH} \approx 875 \msun$. Globally, the azimuthally averaged gas density falls approximately as $\rho \propto r^{-2}$, but the full density distribution is produced by giving the gas a turbulent velocity field (following the method outlined in \citealt{Dubinski1995ApJ} and \citealt{Hobbs2011MNRAS}) and allowing the gas shell to relax for $1$~Myr. The initial characteristic turbulent velocity is $\sigma_{\rm t} = 149$~km~s$^{-1}$.

\begin{table}
  \centering
  \def\arraystretch{1.2}
  \caption{Summary of simulations, including key results.}
  \begin{tabular}{@{}lccc@{}}    
    \toprule
    Run & $L_{\rm AGN}$ & min$\,M_{\rm in}\left(<0.5\,{\rm kpc}\right)$ & $\Delta M_{\rm SMBH}\left(1\,{\rm Myr}\right)$ \\  
    & $\left(\text{erg s}^{-1}\right)$ & $\left(\msun\right)$ & $\left(\msun\right)$ \\
    \midrule
    Control & $0$                & $1.5 \times 10^{8*}$& $1.2 \times 10^{8}$ \\
    L0.1 & $1.26 \times 10^{45}$ & $1.5 \times 10^{8*}$& $1.2 \times 10^{8}$ \\
    L0.3 & $3.78 \times 10^{45}$ & $1.4 \times 10^{8}$   & $4.0 \times 10^{7}$ \\
    L0.5 & $6.30 \times 10^{45}$ & $9.2 \times 10^{7}$   &  $5.1 \times 10^{6}$ \\
    L0.7 & $8.82 \times 10^{45}$ & $2.8 \times 10^{7}$   &  $1.6 \times 10^{6}$ \\
    L1.0 & $1.26 \times 10^{46}$ & $7.8 \times 10^{6}$   &  $6.2 \times 10^{4}$ \\
    L1.3 & $1.64 \times 10^{46}$ & $4.1 \times 10^{6}$   &  $5.9 \times 10^{3}$ \\
    L1.5 & $1.89 \times 10^{46}$ & $2.3 \times 10^{6}$   &  $4.4 \times 10^{3}$ \\
    L1.7 & $2.14 \times 10^{46}$ & $1.3 \times 10^{6}$   &  $6.1 \times 10^{3}$ \\
    L1.9 & $2.39 \times 10^{46}$ & $6.0 \times 10^{5}$   &  $3.9 \times 10^{3}$ \\
    L2.0 & $2.52 \times 10^{46}$ & $3.1 \times 10^{5}$   &  $4.8 \times 10^{3}$ \\
    L2.2 & $2.77 \times 10^{46}$ & $2.3 \times 10^{5}$   &  $2.8 \times 10^{3}$ \\
    L2.5 & $3.15 \times 10^{46}$ & $8.1 \times 10^{4}$   &  $5.7 \times 10^{3}$ \\
    \hline
  \end{tabular}
  \begin{minipage}{\columnwidth}
  \vspace{0.2cm}
    \small
    \textbf{Notes:} The first column shows the simulation name in the form L\#, where the number refers to AGN luminosity in terms of the Eddington luminosity $L_{\rm Edd} = 1.26\times10^{46}$~erg~s$^{-1}$. The second column shows the AGN luminosity. The subsequent columns show the main quantitative results: minimum value of cold rapidly infalling gas mass within the central $0.5$~kpc and the SMBH particle mass change over $1$~Myr. Each line corresponds to four simulations with stochastically different initial conditions, which we used to investigate the level of variance due to small-scale inhomogeneities. $^*$ marks models where the minimum is at the beginning of the simulation run (see the main text for details).    
  \end{minipage}
  \label{tab:sims}
\end{table}

As the simulation proceeds, the gas is affected by external gravity (we neglected self-gravity in the simulations; the importance of this assumption is discussed in Sect. \ref{sec:self_gravity}) and AGN wind feedback. The wind was not modelled hydrodynamically, but instead propagated on a static spherical grid, as described in Sect. 3.3 of Paper I. This prescription ensures a spherically symmetric injection of AGN wind energy and momentum and conserves these quantities, as we show in Appendix \ref{sec:injection}. We employed two sub-resolution prescriptions for gas cooling. At temperatures $T_{\rm g} > 10^4$~K, we used the prescription of \citet{Sazonov2005MNRAS}, which models the heating and cooling of optically thin gas due to a typical AGN radiation field. It includes photoionisation heating, the Compton effect and metal line cooling. We modified this function by neglecting the effect of Compton cooling, a change appropriate for the expected two-temperature plasma nature of the hottest outflowing gas \citep[see][]{Faucher2012MNRASb, Bourne2013MNRAS}. Overall, the heating effect was negligible, especially for dense gas, as expected \citep[cf.][who showed that direct thermal feedback is only relevant for gas with densities 1-2 orders of magnitude lower than the average density in our simulations]{Sazonov2005MNRAS}. Below $T_{\rm g} = 10^4$~K, the gas cooled according to the prescription of \citet{Mashchenko2008Sci},  which approximates the cooling effect of metastable C, N, O, Fe, S, and Si lines in ionisation equilibrium maintained by locally produced cosmic rays at approximately Solar metallicity. This function allowed the gas to cool to $20$~K. All gas particles that fell closer than $r_{\rm acc} = 0.01$~kpc from the black hole particle were removed from the simulation. We used this to track the expected accretion onto the SMBH, but did not adjust the AGN luminosity because the time for gas to fall from $\sim 10$~pc through the accretion disc to the SMBH is comparable to the duration of our simulations. A summary of the main simulation parameters and salient results is given in Table \ref{tab:sims}.

\section{Results} \label{sec:results}

\begin{figure*}
\includegraphics[width=1\textwidth]{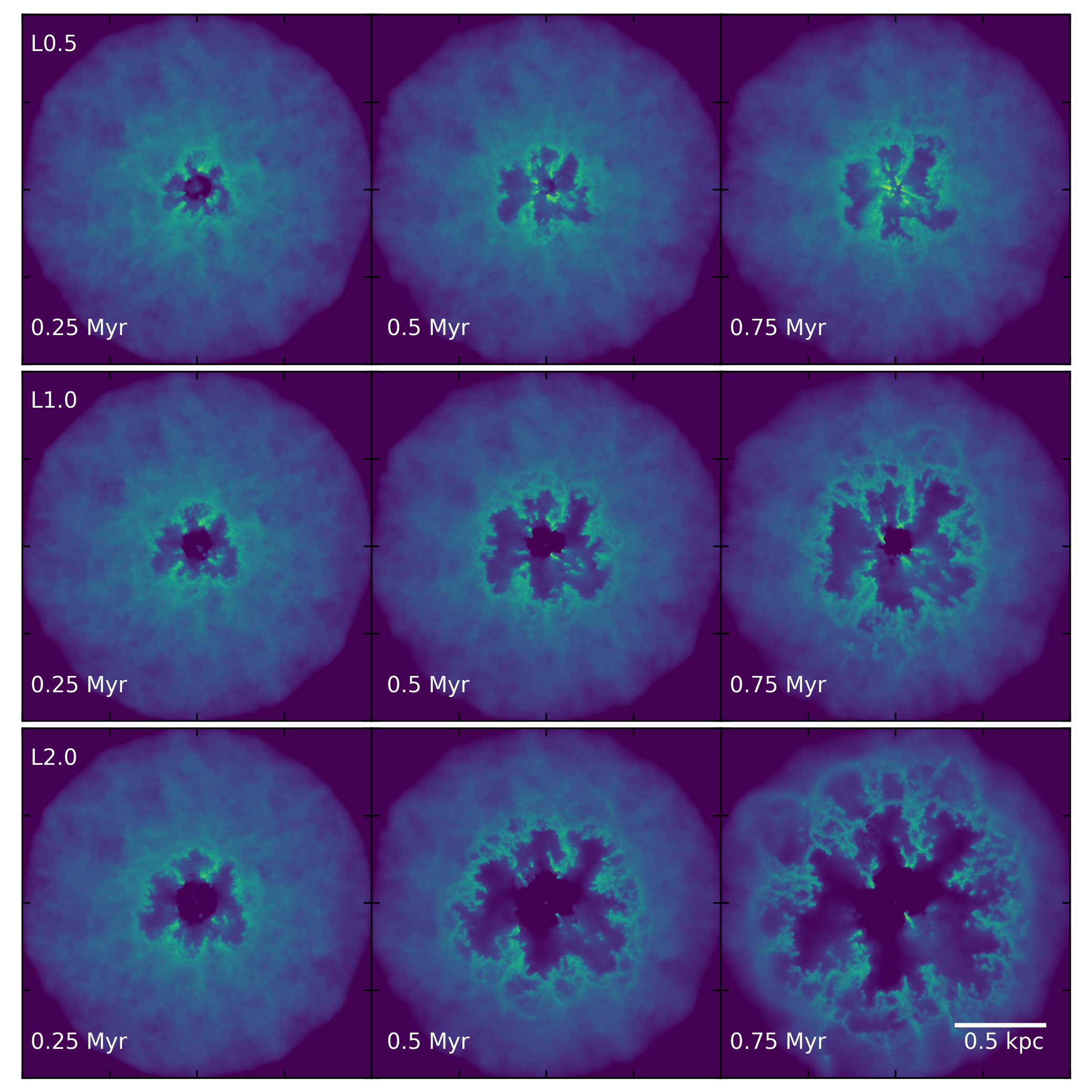}
\caption{Gas density integrated through a $20^\circ$ wedge around the mid-plane ($z=0$), defined by $\left|z\right| < 0.18\left(x^2+y^2\right)$, in simulations L0.3 (top row), L1.0 (middle), and L1.7 (bottom). Columns show $t=0.5$, $0.75$, and $1.0$ Myr from left to right. Brighter colours indicate higher gas densities, in the range $\sim10^{-25}$~g~cm$^{-3} < \rho <10^{-19}$~g~cm$^{-3}$.
}
\label{fig:all_morphology}
\end{figure*}

Qualitatively, the simulations can be divided into three groups. When $L_{\rm AGN}/L_{\rm Edd} \leq 0.5$, the AGN is too weak to significantly affect gas infall. When $0.7 \leq L_{\rm AGN}/L_{\rm Edd} \leq 1.7$, the effect is moderate and allows some infall, although most of the gas is pushed away. AGNs with $L_{\rm AGN}/L_{\rm Edd} \geq 1.9$ essentially completely shut down gas infall. We therefore mostly focus on presenting the results of three simulations --- L0.5, L1.0 and L2.0 --- in detail, including the control simulation where appropriate, and limit the depiction of other simulations' results to those plots where we show direct parameter dependence on AGN luminosity.

We start by showing density maps to describe the qualitative evolution of the gas distribution. We then analyse the properties of infalling gas, using kinematic phase maps, radial mass profiles and time evolution of infalling gas mass to highlight the differences among the three simulations. Then we consider the feedback effect on individual dense gas clusters. Finally, we show the differences in SMBH particle growth and determine the AGN luminosity that would produce a self-consistent growth rate.

\subsection{Gas morphology}

\begin{figure*}
\includegraphics[width=0.32\textwidth, trim={0 0 0 0}, clip]{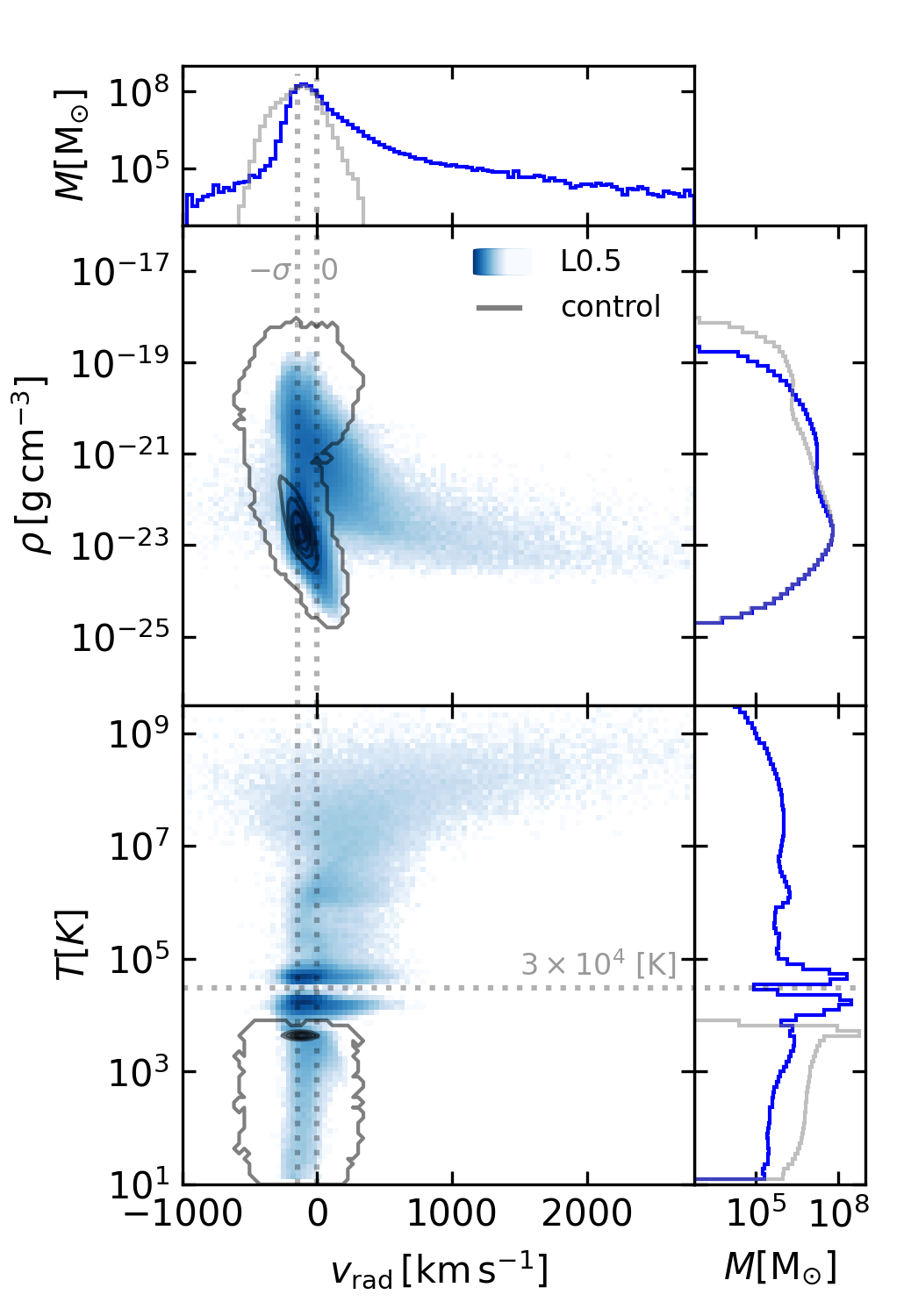}
\includegraphics[width=0.32\textwidth, trim={0 0 0 0}, clip]{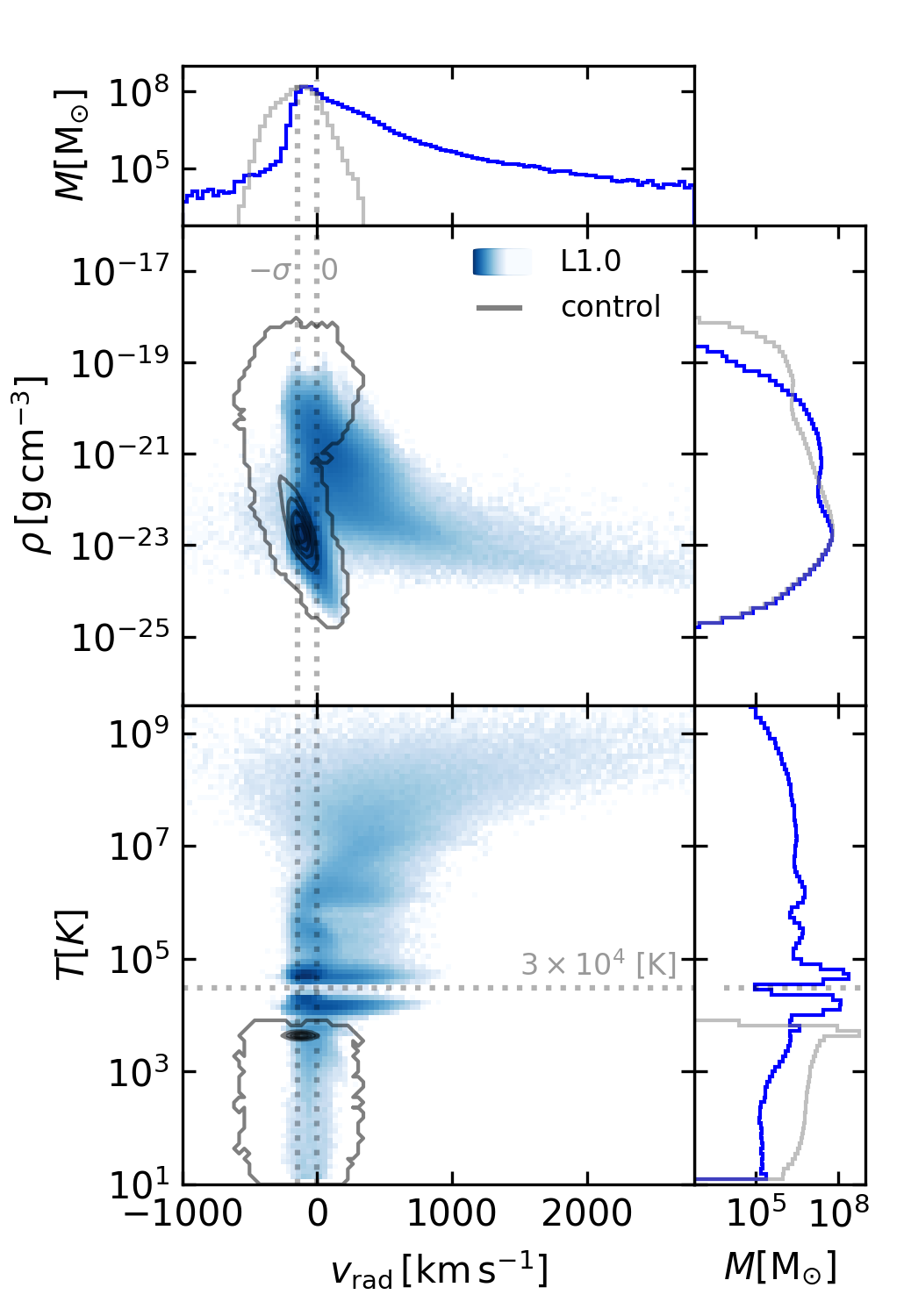}
\includegraphics[width=0.32\textwidth, trim={0 0 0 0}, clip]{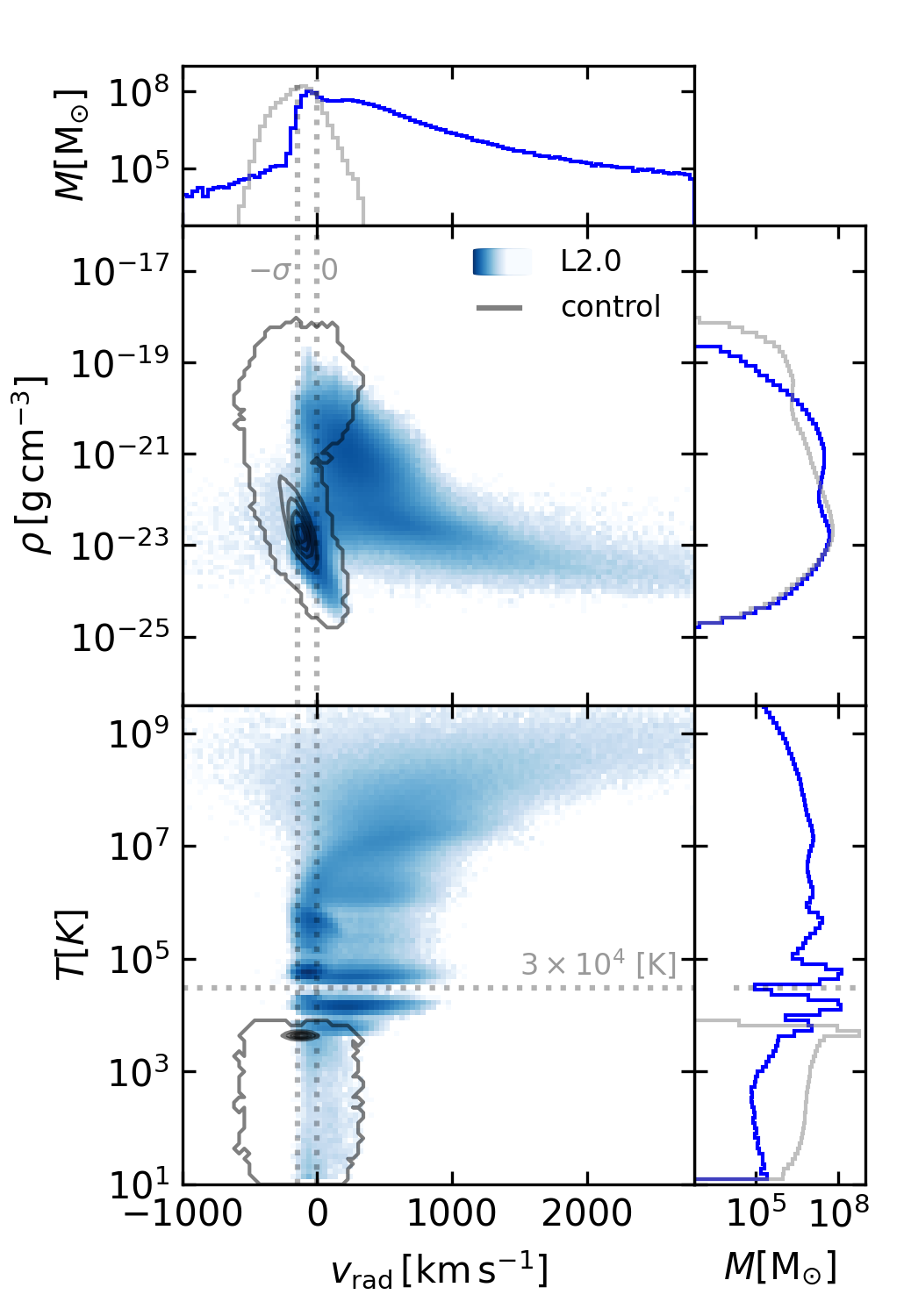}

\caption{Kinematic phase diagrams at $t=0.5$~Myr in simulations L0.5 (left), L1.0 (middle), and L2.0 (right). In each panel, the top 2D histogram shows radial velocity against density, and the bottom the radial velocity against temperature. 1D histograms at the top and right show distributions of individual gas properties. Grey contours show corresponding distributions in the control simulation. Vertical dotted lines show $v_{\rm r} = 0$ and $v_{\rm r} = -\sigma_{\rm b}$; horizontal dotted line shows $T = 3\times10^4$~K.}
\label{fig:velophase}
\end{figure*}

Figure~\ref{fig:all_morphology} shows the evolution of gas density in runs L0.5, L1.0, and L2.0 over $0.25$–$0.75$ Myr. In L0.5, the AGN disturbs the gas and inflates bubbles that gradually expand to a radius $\sim 0.3$~kpc in all directions by $0.75$~Myr, although gas disturbances are seen in some directions as far as $\sim 0.5$~kpc. This corresponds to an average velocity $\sim 400-670$~km~s$^{-1}$, somewhat lower than the analytical expectation $v_{\rm e} \approx 680$~km~s$^{-1}$ (Eq. \ref{eq:dotr_en}). However, the bubbles are faint and only affect the diffuse gas. The region around the SMBH is not cleared out; dense gas continuously falls inwards and feeds the SMBH particle, as we show below.

In simulation L1.0 (middle row), the outflow is significantly faster ($v_{\rm r} \sim 670-1000$~km~s$^{-1}$) and almost entirely clears the central $\sim 100$~pc by 1~Myr. SMBH accretion is essentially shut down during the activity phase. However, dense gas `fingers' remain and even approach the nucleus, so once the AGN switches off, they may provide significant fuel to restart the activity and continue growing the SMBH. 

Finally, in simulation L2.0, the outflow breaks out of the initial gas shell by $t=0.75$~Myr. The effect on dense gas is particularly noticeable: the outflow clears a $\sim 200$~pc-wide region by $0.75$~Myr and pushes out both dense and more diffuse gas. Additionally, dense gas is confined to isolated elongated clouds rather than `fingers' as in L1.0.

\subsection{Mass infall}

\begin{figure}
\includegraphics[width=\columnwidth]{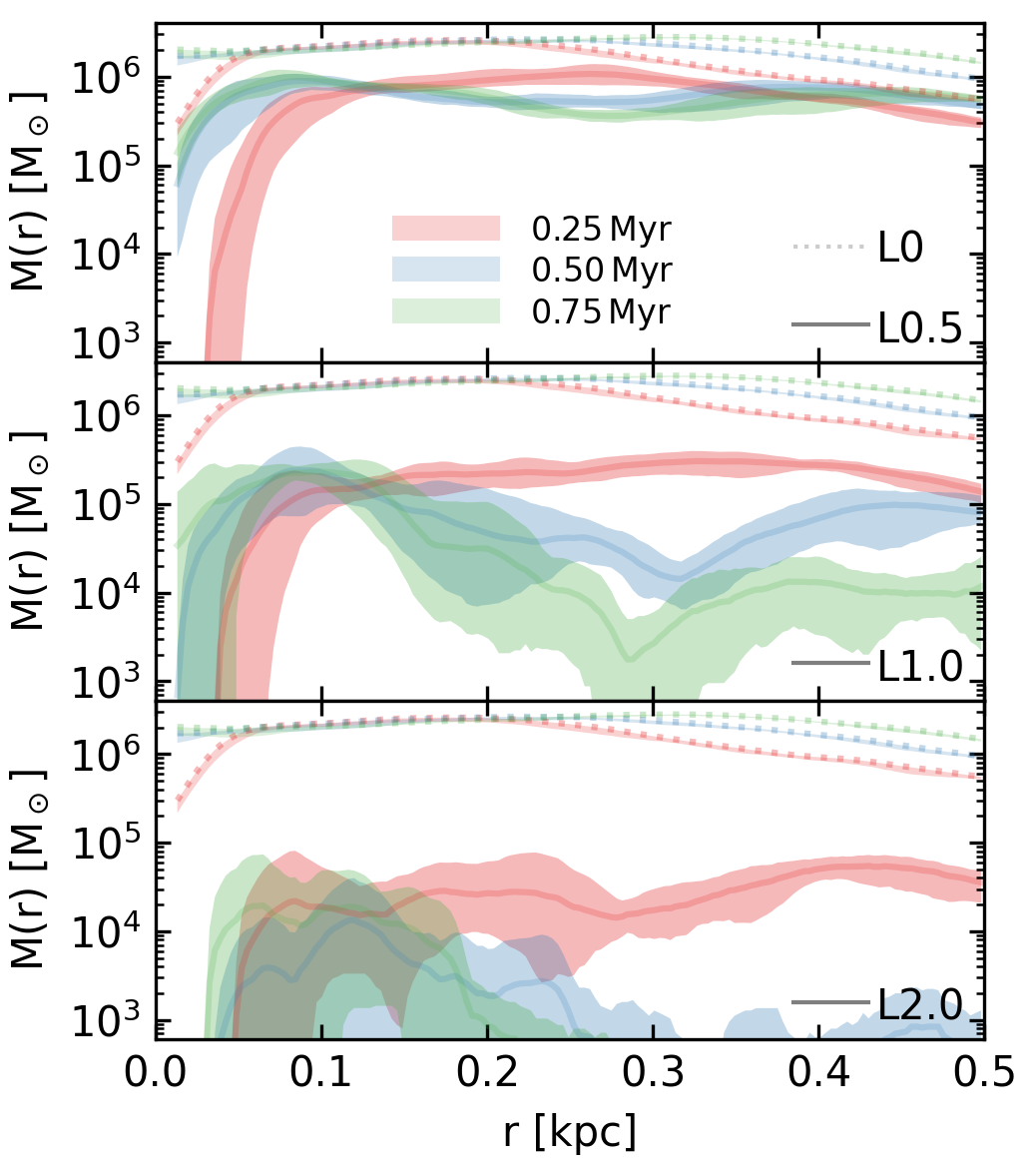}
\caption{Mass of cold rapidly infalling gas in radial $3.3$-pc-wide bins at $0.25$, $0.5,$ and $0.75$~Myr (solid red, blue, and green lines and shading, respectively) in simulations L0.5, L1.0, and L2.0 (top, middle, and bottom panels, respectively). Solid lines show the mean of the four stochastically different simulations; shading encompasses their full range. Dotted lines in each panel show equivalent results from the control simulation.}
\label{fig:infalling_profile}
\end{figure}

\begin{figure}
\includegraphics[width=0.49\textwidth, trim={0 0 0 0}, clip]{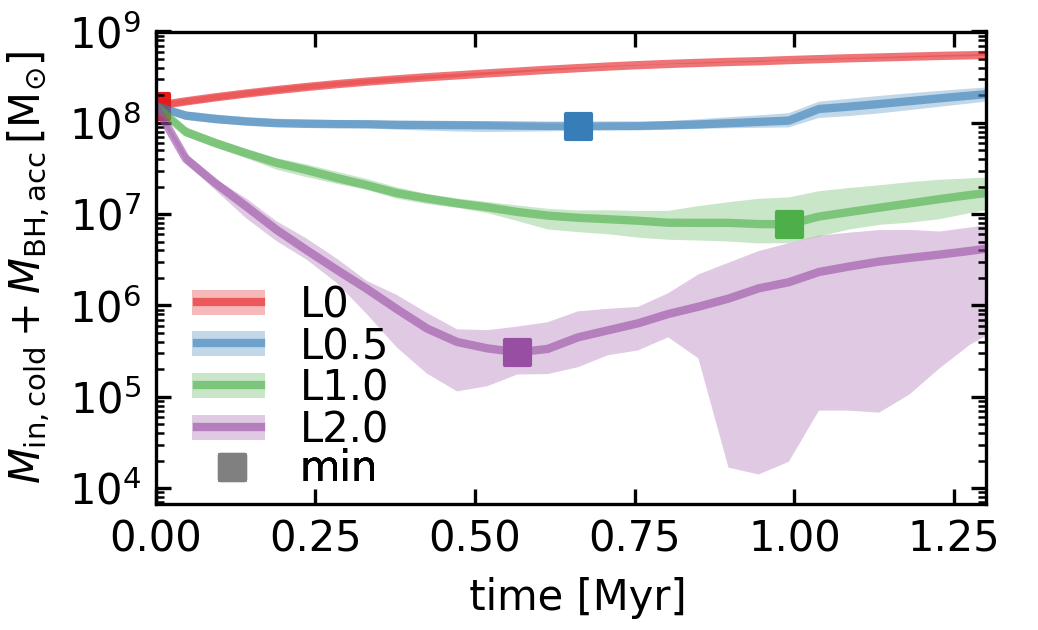}
\includegraphics[width=0.49\textwidth, trim={0 0 0 0}, clip]{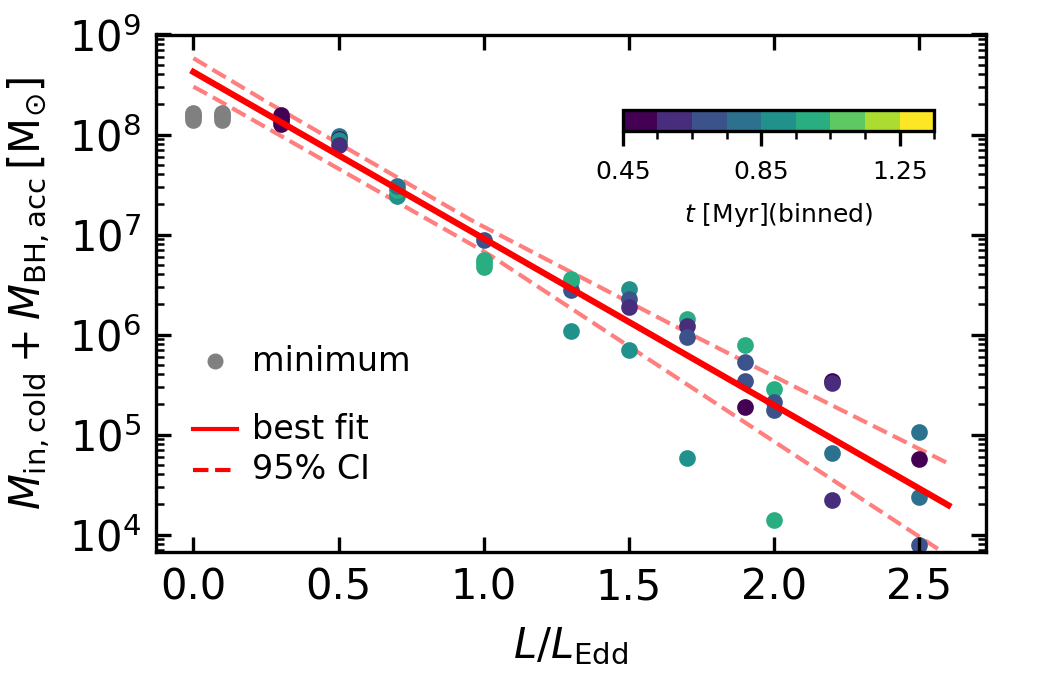}
\caption{ Top: Total mass of cold rapidly infalling gas within the central $500$~pc as a function of time in the control (red), L0.5 (blue), L1.0 (green), and L2.0 (magenta) simulations. We added the mass accreted by the SMBH particle to this total. Solid lines show the mean of the four stochastically different simulations; shading encompasses their full range. Squares mark the times when the mean mass reaches its minimum value. Bottom: Minimum value of cold rapidly infalling gas mass in all simulations (circles) as a function of luminosity, with colours denoting the time when the minimum is reached. The solid red line is the best linear fit between $\log{M}$ and $L_{\rm AGN}$ of all simulations except the control and L0.1; the dashed lines indicate $95\%$ confidence limits on the line parameters, estimated via bootstrapping.}
\label{fig:infalling}
\end{figure}

To get a better sense of the properties of infalling gas, we show in Fig.~\ref{fig:velophase} kinematic phase diagrams across the three luminosity runs together with our adopted criterion for selecting `cold' ($T < 3\times 10^4$~K) and `rapidly infalling' ($v_{\rm rad} \leq -\sigma_{\rm b}$) gas. The latter criterion identifies gas with inward motion that is not the result of turbulence. All distributions are qualitatively similar, with a very high phase density region corresponding to the outer parts of the initial distribution that have not been affected by the outflow yet. Gas radial velocities range from $\sim -1000$~km~s$^{-1}$ to $> 4000$~km~s$^{-1}$, although most of the gas has velocities much closer to zero. Some gas has more negative radial velocities than the lowest values in the control simulation. Its low density and high temperature indicate hot gas expanding into the region evacuated by the outflow. This effect may be a numerical artefact caused by the lack of hydrodynamical particles representing the AGN wind \citep[see][]{Zubovas2024AA}. The gas density distribution does not change very significantly compared with the control model, except for the maximum gas density decreasing by a factor of $\sim 1.5-2$ in all simulations. This suggests that AGN feedback prevents the formation of the densest clumps. Nevertheless, a lot of dense gas with $\rho > 10^{-21}$~g~cm$^{-3} (n \gtrsim 10^3$~cm$^{-3})$ exists in all simulations. The temperature distribution is also qualitatively similar in all simulations with AGNs, but differ markedly from the control model, because even a low-luminosity AGN can rapidly heat and accelerate diffuse gas. The multiple peaks seen in the temperature histogram are imposed by the adopted cooling function, which has minima at $T \sim 10^4$~K and $T \sim 5\times 10^4$~K. The highest temperatures result from shock heating of the outflowing gas. We used the approximate midpoint between the two peaks, $T = 3\times10^4$~K, to divide the gas into `cold' and `warm-hot' phases. Subsequently, we focused on the infall of cold gas, which can be gravitationally bound to the SMBH. 

For the purposes of our investigation, the most significant result is the change of dense gas ($\rho > 10^{-21}$~g~cm$^{-3}$) velocities with $L_{\rm AGN}$. This gas is exclusively cold and lacks thermal energy to escape the SMBH gravitational potential. As a result, it can feed the SMBH efficiently when inflowing. In the control simulation, most of the dense gas is infalling by $t = 0.5$~Myr, as expected, as the gravitational potential overcomes decaying turbulent velocities. In L0.5, the balance shifts towards higher velocities with essentially no gas having $v_{\rm rad} < -280$~km~s$^{-1}$. However, most dense gas retains negative radial velocities. Higher AGN luminosities gradually change this distribution: in L1.0, the lowest velocities of infalling dense gas at $t = 0.5$~Myr increase to $v_{\rm rad} \sim -180$~km~s$^{-1}$, while in L2.0, they further increase to $v_{\rm rad} \sim -145$~km~s$^{-1}$. Additionally, in L1.0, the total mass of infalling dense gas becomes comparable to the outflowing mass, while in L2.0, the outflow clearly dominates.

The radial gas distribution (Fig. \ref{fig:infalling_profile}) also reveals the evolution of outflowing and inflowing material. In L0.5, the outflow is initially able to clear a $\sim 30$~pc region (also seen in Fig. \ref{fig:all_morphology}, top left). However, the infalling gas quickly returns and fills the central region, reaching $(0.7-1)\times10^6\,\msun$ per radial bin, approximately half the value in the control simulation. In L1.0, the situation is markedly different. Initially, the outflow clears the central $\sim 40$~pc. Beyond that, the gas mass decreases, but persistent dense clumps remain at $r \lesssim 250$~pc after the outflow has passed. Some of them gradually approach the SMBH, reaching the sink radius around $t = 0.5$~Myr. Outside this region, cold gas is gradually evacuated and does not return until the end of the simulation. The total infalling mass remaining in the central region is only a few per cent of the control simulation value. In L2.0, evacuation is more efficient: the bulk of the material is cleared within $t\approx0.5$~Myr, and the outflow maintains a clear zone of $r \sim 50$~pc until the end of the simulation. The total mass of cold rapidly infalling gas in each bin never exceeds $1\%$ of that in the control simulation.

The top panel of Fig. \ref{fig:infalling} shows the time evolution of rapidly infalling cold gas mass within the central $500$~pc in the four simulations, including control. 
We interpret these values as the maximum gas mass that could feed the SMBH once the AGN episode has finished (neglecting possible star formation, discussed in Sect. \ref{sec:self_gravity}). The total includes the gas mass accreted by the SMBH particle, though its contribution remains below $10\%$ except in the control simulation.
In the control simulation, the infalling gas mass increases continuously as gas from outer regions approaches the SMBH. In L0.5, the outflow delays this process, but cannot prevent cold dense clumps from falling inwards, maintaining a large total infalling mass. L1.0 shows a gradual decrease from an initial value just above $10^8 \, \msun$ to a minimum of $\sim 7.8\times10^6\,\msun$ around $t=1$~Myr. Later, when dense clumps overtaken by the outflow fall inwards, the infalling gas mass slowly increases to $\sim 2\times10^7 \, \msun$. Note that the minimum mass is reached at significantly different times in stochastically different simulations: two reach minima already at $t \sim 0.75$~Myr; the minimum mass varies by a factor of $\sim 3$. In L2.0, the decrease is faster, down to $\sim 3.1\times10^5\,\msun$ by $t=0.6$~Myr, but later the infalling mass increases back up to $\sim 4\times10^6 \, \msun$ for the same reasons as L1.0. The time of minimum value shows a similar scatter among stochastically different simulations as L1.0, but the minimum mass ranges over more than two orders of magnitude.

The bottom panel shows the minimum rapidly infalling cold gas mass within the central $500$~pc in each simulation versus AGN luminosity. 
In the control and L0.1 simulations, the infalling mass only increases, so the minimum occurs at $t = 0$ and is equal to the initial cold rapidly infalling gas mass $M_{\rm cold,in,init}$. With increasing luminosity, the minimum mass follows an exponential trend: $M_{\rm cold,in,min}\approx 4.3\times10^8 \exp\left(-1.67L/L_{\rm Edd}\right)\,\msun$. Note that $M_{\rm cold,in,min} < 0.1\times M_{\rm cold,in,init}$ in all simulations with $L \geq L_{\rm Edd}$ and lower than $M_{\rm cold,in,min} < 0.01\times M_{\rm cold,in,init}$ when $L\geq 1.9 L_{\rm Edd}$. We conclude that AGNs with luminosities $L \gtrsim L_{\rm Edd}\left(M_\sigma\right)$ can significantly disrupt and remove dense gas reservoirs that would otherwise feed the SMBH.

\subsection{Feedback on dense gas clumps}

\begin{figure}
\centering
\hspace{0.025\textwidth}\includegraphics[width=0.45\textwidth, trim={0 0 0 0}, clip]{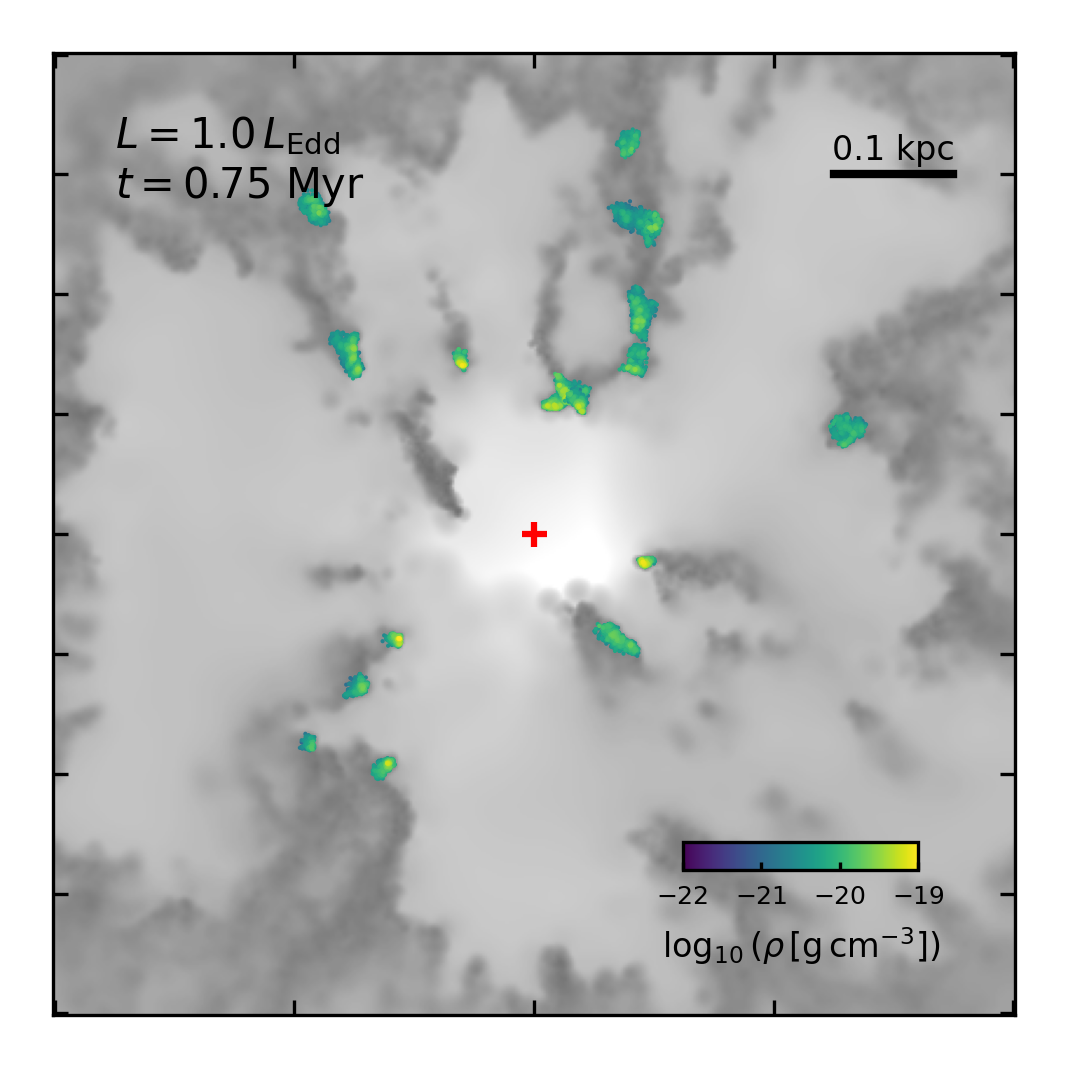}
\includegraphics[width=0.49\textwidth, trim={0 0 0 0.5cm}, clip]{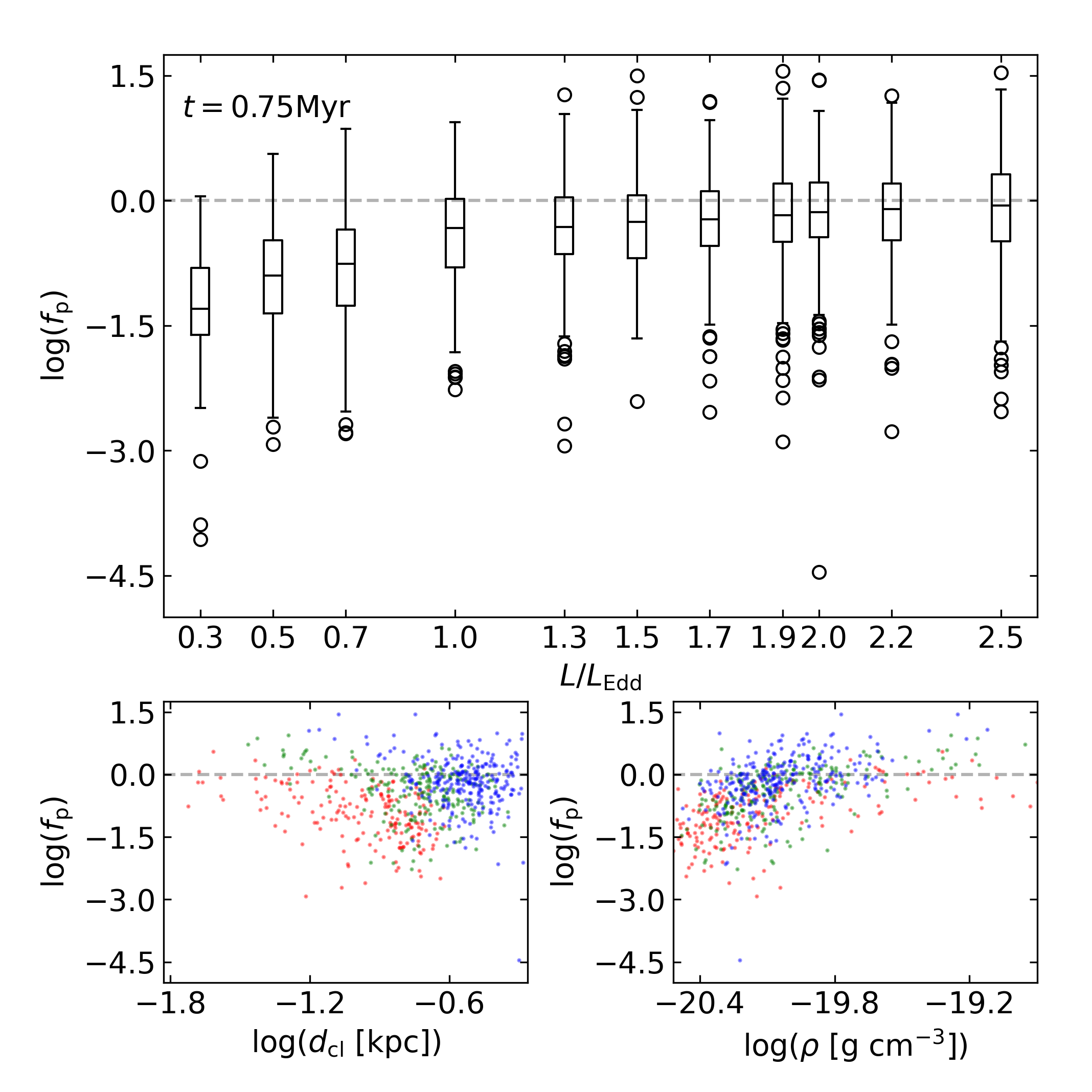}
\caption{Top: Clusters (coloured by gas density) detected at $t = 0.75$~Myr in simulation L1.0, plotted over the gas density map (grey scale). Middle panel: Momentum loading factor ($f_{\rm p}$) distributions (Eq. \ref{eq:fp}; box and whisker plots) in simulations with different AGN luminosities. Circles highlight outliers beyond 1.5 times the interquartile range covered by the whiskers. The horizontal dashed line highlights $f_{\rm p} = 1$, which corresponds to pure momentum driving, for ease of comparison. Bottom panels: Momentum loading factor ($f_{\rm p}$) as a function of $d_{\rm cl}$ (left) and mean gas density (right) in simulations L0.5 (red), L1.0 (green), and L2.0 (blue).}
\label{fig:clusters}
\end{figure}

Each gas particle experiences feedback differently not only because of its properties, such as density or temperature, but also due to environment. For example, diffuse gas can be shielded by dense clumps, while particles can enter and leave such clumps. To quantify the effect of AGN feedback on cold dense gas, we identified clusters of particles at $t=0.75$~Myr and tracked their evolution between two snapshots, i.e. over $\Delta t = 0.05$~Myr. Using DBSCAN from sklearn \citep{Ester1996, Pedregosa2011}, we searched for clusters defined as groups of at least 100 particles with density $\rho > 10^{-22}$~g~cm$^{-3}$ and $T < 3\times10^4$~K, whose particles have nearest neighbours closer than $\delta d < 10$~pc and relative velocities $\delta \vec{v} < 0.5\sigma$. The top panel of Fig. \ref{fig:clusters} shows detected clusters in simulation L1.0 superimposed on the grey scale density map. Not all dense regions are assigned to clusters, usually because they contain fewer than 100 particles.

We quantified the AGN wind effect on each cluster using Eq. \ref{eq:eom_momentum} to calculate the momentum loading factor:
\begin{equation}\label{eq:fp}
    f_{\rm p} = \frac{\Delta\left(M_{\rm cl}v_{\rm rad,cl}\right)/\Delta t + G M_{\rm cl}M_{\rm pot}\left(<d_{\rm cl}\right)/d_{\rm cl}^2}{L_{\rm AGN}/c \times \Omega_{\rm cl}/4\pi},
\end{equation}
where $M_{\rm cl}$ is the cluster mass, $v_{\rm rad,cl}$ is its mean radial velocity, $d_{\rm cl}$ is the distance between the origin and its centre of mass and $\Omega_{\rm cl}$ is the solid angle subtended by the cluster when viewed from the origin. The numerator contains the change in cluster radial momentum over time and the gravitational force from the background potential; the denominator is the fraction of the AGN wind momentum $L_{\rm AGN}/c$ directly impinging on the cluster.

In the middle panel of Fig. \ref{fig:clusters}, we show the distributions of $f_{\rm p}$ in simulations with different AGN luminosities at $t = 0.75$~Myr using box-and-whisker plots. A spherically symmetric energy-driven outflow has $f_{\rm p} \sim v_{\rm w} / v_{\rm out} \gtrsim 30$, higher than almost all clusters. Despite the large scatter, most values fall in the range $0.1 < f_{\rm p} < 1.4$, with a mean of $f_{\rm p} \approx 0.33$ and a slight positive correlation with luminosity. So most clusters experience an effective force lower than or comparable to that of the AGN wind momentum, even though the shocked wind does not cool. This results from self-shielding: gas on the inner edge of the cluster is exposed to the wind and absorbs its momentum, while gas farther away is shielded even from this modest effect. Values of $f_{\rm p}>1$ occur when clusters are pushed by the shocked wind plasma before it escapes through low-density channels.

In the bottom two panels of Fig. \ref{fig:clusters}, we plot $f_{\rm p}$ versus $d_{\rm cl}$ and mean cluster gas density. 
Each simulation shows a slight negative correlation between $f_{\rm p}$ and $d_{\rm cl}$. This results from the shocked AGN wind confinement: the hot plasma attempts to escape through low-density channels, but cannot do so indefinitely fast and pushes against dense gas too. Near the SMBH, clusters subtend relatively large solid angles, so the shocked wind plasma takes longer to escape and pushes them more. There is a positive correlation between $f_{\rm p}$ and cluster density. In denser clusters, self-shielding is less important because the material on the inner edge of the cluster efficiently pushes against that further out. We find no significant correlations with other cluster parameters, such as their sizes, temperatures, masses or radial velocities.

These results show that cold dense gas clusters experience almost exclusively momentum feedback from the AGN wind, even while the shocked wind plasma remains hot and drives an energy-conserving outflow of diffuse gas.

\subsection{SMBH growth}

\begin{figure}

\includegraphics[width=0.49\textwidth, trim={0 0 0 0}, clip]{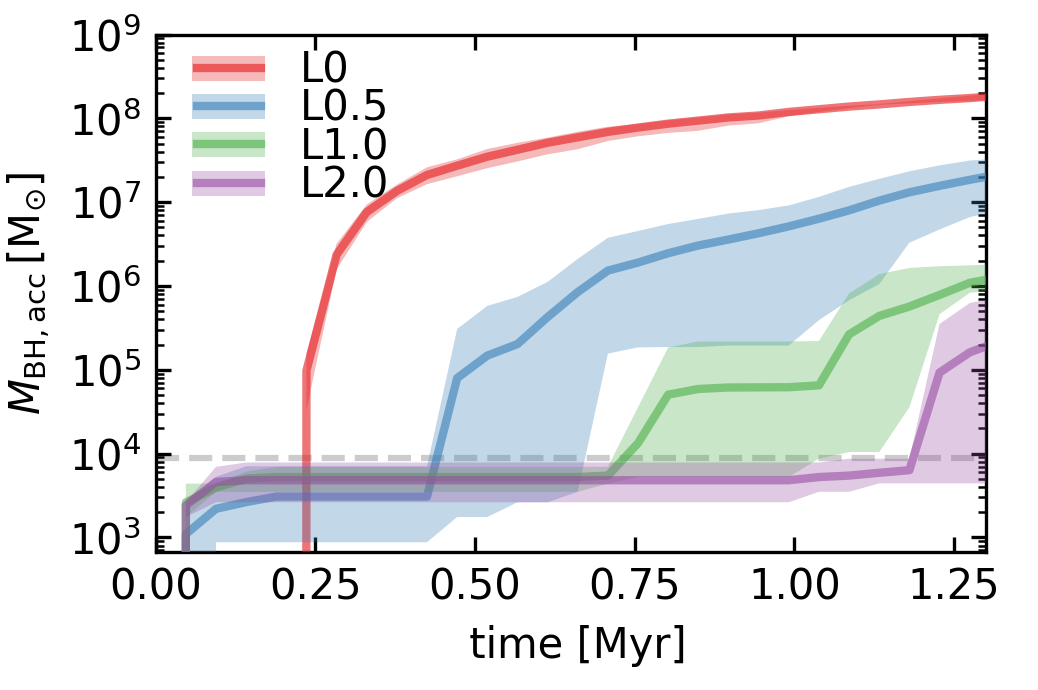}
\includegraphics[width=0.49\textwidth, trim={0 0 0 0}, clip]{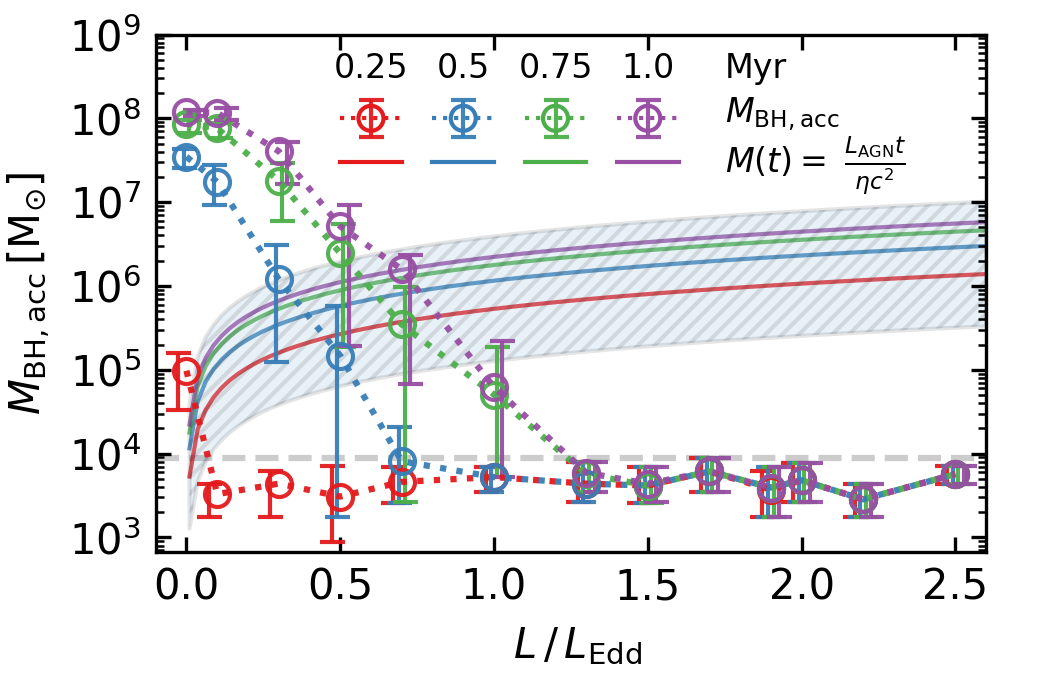}
\caption{Top: Same as the top panel of Fig. \ref{fig:infalling} but showing the change in SMBH particle mass. The horizontal dashed line shows $M = 10 m_{\rm SPH}$. Bottom: Change in SMBH particle mass between the start of the simulation and $t = 0.25$, $0.5$, $0.75,$ and $1$~Myr (red, blue, green, and magenta circles, respectively), in simulations with different AGN luminosities. Error bars represent the range of values in stochastically different simulations; they are slightly offset horizontally for clarity. Solid lines show the SMBH mass change expected due to accretion powering an AGN of the given Eddington fraction over the same time intervals, assuming a radiative efficiency ($\eta$) of $0.1$; the shaded region represents the effect of varying radiative efficiency between $0.057$ and $0.42$. The horizontal dashed line is the same as in top panel.}
\label{fig:smbh_mass}
\end{figure}

Despite AGN outflow feedback, some gas falls through the sink radius and is accreted by the SMBH particle. We show this Fig. \ref{fig:smbh_mass}. In the top panel, we plot the evolution of total mass accreted by the sink particle. In all simulations except control, the SMBH rapidly consumes a few particles ($<10^4 \, \msun$) heated by the AGN, followed by a quiescent period of zero growth, when the outflow has cleared the inner region and dense gas clumps have not returned. In L0.5, this period lasts around $0.4$~Myr; later, the SMBH particle begins growing at an accelerating rate, consuming $\sim 5\%$ of its initial mass within $<1$~Myr, equivalent to an average accretion rate $\sim 2.5$ times higher than Eddington. However, the SMBH itself would not grow at this rate, because the SMBH particle growth represents infall of gravitationally bound matter through the sink radius at $r_{\rm sink} = 10$~pc. In L1.0, the quiescent period lasts $0.75$~Myr, followed by two short yet significant bursts of accretion, swallowing $\sim 6\times10^4\,\msun$ in $\sim 0.1$~Myr and another $\sim 10^6 \, \msun$ in $0.2$~Myr about $0.25$~Myr later. These bursts correspond to accretion rates $\sim 0.3-2.5$ times Eddington. Nevertheless, the SMBH particle growth is significantly limited by the available mass supply, which in this simulation is only $\sim 7.8\times10^6\,\msun \sim 0.08 M_{\rm BH}$. The SMBH cannot grow by more than a few per cent of its initial mass because the AGN episode has expelled most of the surrounding gas. In L2.0, the quiescent period lasts $\sim 1.2$~Myr, followed by only one short accretion burst of a few times $10^5\,\msun$. Based on the evolution of radial gas profiles (see the previous subsection), we expect at most a few more such bursts powered by the remaining available gas.

The behaviour described above suggests that an AGN shining at $L = 0.5 L_{\rm Edd}$ is unable to prevent Eddington-level accretion onto the SMBH, while a luminosity exceeding $L = L_{\rm Edd}$ is sufficient for this. To quantify this trend, we show, in the bottom panel of Fig. \ref{fig:smbh_mass}, the SMBH particle mass change against AGN luminosity and the mass change that would result from accretion producing that luminosity. It is obvious that in simulations with $L \leq 0.5 L_{\rm Edd}$, the SMBH particle accretes more gas that would be needed to maintain the luminosity, i.e. feedback is unable to regulate the accretion rate. Conversely, when $L_{\rm AGN} \gtrsim 0.7 L_{\rm Edd}\left(M_\sigma\right)$, the AGN severely limits the accretion rate on to the SMBH, i.e. feedback is effective at preventing further SMBH growth. After an AGN episode with $L_{\rm AGN} \gtrsim 0.7 L_{\rm Edd}\left(M_\sigma\right)$, the SMBH would stay quiescent for a prolonged period, potentially indefinitely, unless external factors (e.g. instabilities, galaxy mergers, tidal disruption events) reignite the AGN later. So the critical AGN luminosity required for feedback to become efficient in regulating SMBH growth is very close to the Eddington luminosity for a black hole with mass given by $M_\sigma$ (i.e. Eq. \ref{eq:msigma}). This fact strongly supports the hypothesis that momentum feedback is responsible for establishing the $M-\sigma$ relation, even when the shocked AGN wind transfers all of its kinetic power to the surrounding gas.

\section{Discussion} \label{sec:discuss}

\subsection{Establishing and maintaining the $M-\sigma$ relation}

We have shown that an energy-driven outflow expanding through a turbulent gas distribution affects cold dense gas primarily through momentum injection. This, in principle, recovers the critical luminosity condition, based on the momentum-driven outflow paradigm, for shutting down further SMBH growth \citep{Murray2005ApJ}, which leads to the establishment of the $M-\sigma$ relation \citep{King2010MNRASa}. Below, we outline  several additional steps between our results and the establishment of the relation that our simulations are not designed to track.

Once the dense clumps and filaments are removed, further accretion of material on to the SMBH must be prevented or, at least, its average rate reduced significantly. In simulations with high AGN luminosities, $L > 0.7 L_{\rm Edd}$, the amount of dense infalling gas is reduced to $<0.1 M_{\rm BH}$. However, once the AGN switches off, some of the outflowing gas should inevitably decelerate and fall back on to the SMBH, even if the outflow itself persists for many times longer than the AGN episode \citep{King2011MNRAS, Zubovas2023MNRAS}. This can restart the AGN after as little as $t_{\rm fall} \sim R/\sigma \sim 3.5$~Myr, assuming gas infall from outside the $500$~pc radius. The duty cycle implied by this timescale, compared with the duration of our simulations ($t_{\rm AGN} \sim 1-1.5$~Myr), is very large, $\delta_{\rm AGN} \sim t_{\rm AGN}/\left(t_{\rm AGN}+t_{\rm fall}\right)\sim 0.2$. 

In between episodes, some cold clumps disperse and their gas is removed more efficiently during later AGN phases. This leads to a decrease in the duty cycle, until eventually the bulge is essentially empty of cold gas, terminating SMBH growth. We can expect the SMBH mass to grow by a factor of a few over these multiple episodes. Our simulation results imply self-regulation of SMBH growth at around $L = 0.7 L_{\rm Edd}$ (Fig. \ref{fig:smbh_mass}, bottom panel), which is equivalent to an Eddington luminosity for a black hole with $M_{\rm BH} = 0.7 M_\sigma$. Increasing this by a factor of a few leads to a final mass close to, or slightly above, the value given by the $M-\sigma$ relation. The fact that the $M$–$\sigma$ relation is established over multiple AGN episodes is consistent with the conclusion that SMBH mass is a better predictor of star formation quenching than any instantaneous AGN property, as indicated by both simulations and observations \citep{Harrison2017NatAs, Martin2018Natur, Martin-Navarro2021MNRAS, Piotrowska2022MNRAS, Bessiere2024AA}.

The $M-\sigma$ relation arose early in cosmic history; it was already present at $z>6$ \citep{Maiolino2024AA,JuodzbalisMaiolino2025sub}. Since then, subsequent galaxy and SMBH growth maintained it. The kinematic phase plots (Fig. \ref{fig:velophase}) show an anti-correlation between density and velocity in the dense gas. This is qualitatively expected because denser gas has a higher weight and needs more force to be pushed away. It follows that in gas-rich galaxies, the luminosity required to shut down accretion is higher than in gas-poor ones. Given that galaxies at earlier cosmic epochs had higher gas masses \citep{Carilli2013ARAA}, we expect high-redshift galaxies to reach higher SMBH masses before long-term quenching can occur. At later times, when gas is consumed and ejected, the luminosity required to eject new infalling gas clouds decreases. As a result, the SMBH no longer grows significantly, even over gigayear timescales, except during major mergers, which also increase the velocity dispersion. Furthermore, since self-regulation of SMBH feeding occurs at lower luminosities as the Universe evolves, the typical Eddington ratios should decrease with time, as observed \citep{Trump2009ApJ, Willott2010AJ, Mazzucchelli2023AA}.

At the highest redshifts ($z \simgt 6$), dense gas streams falling from intergalactic space \citep{Khandai2012MNRAS, Heintz2024Sci} can be completely resilient to AGN feedback, as any outflows generated expand around them, while the AGN wind momentum is insufficient to stop gas infall and push it away \citep[also see][]{Nayakshin2010MNRAS}. This can lead to very efficient SMBH growth, with duty cycles essentially equal to unity and frequent super-Eddington episodes, again as observed \citep{Wu2022MNRAS, Lupi2024AA, Suh2025NatAs}.

\subsection{Self-gravity and star formation} \label{sec:self_gravity}

\begin{figure}
    \centering
    \includegraphics[width=\columnwidth]{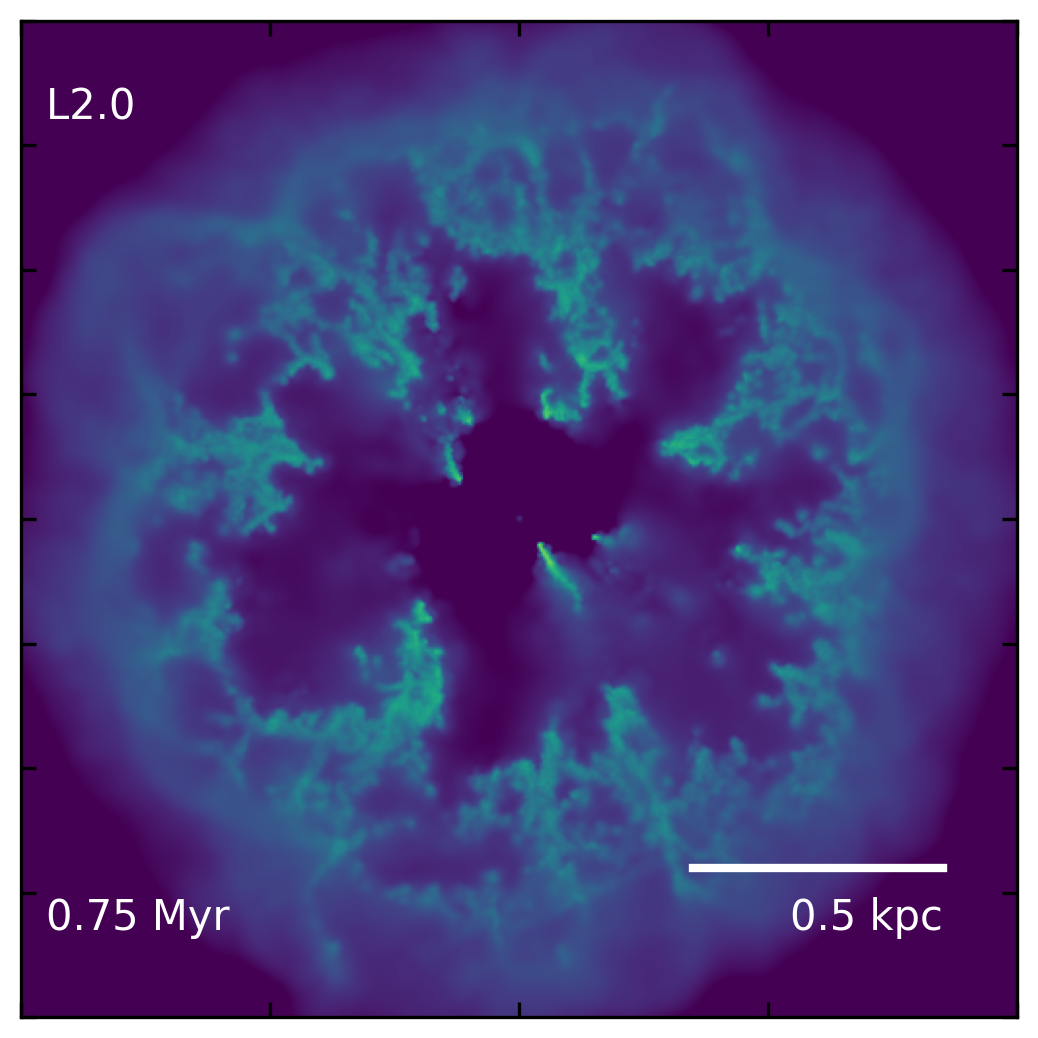}
    \caption{Gas density distribution in simulation L2.0-sg at $t= 0.75$~Myr, directly comparable to the bottom-right panel of Fig. \ref{fig:all_morphology}.}
    \label{fig:selfgravity}
\end{figure}

Our main simulations neglected gas self-gravity and star formation to conserve computational resources. To test the importance of these effects, we set up four additional simulations - L0-sg, L0.5-sg, L1.0-sg, and L2.0-sg. They included gas self-gravity and a crude star formation prescription: gas particles with Jeans mass lower than the resolved mass $m_{\rm res} = 100 m_{\rm SPH} \approx 10^5 \, \msun$ were converted into star particles \citep[see e.g.][for details]{Tartenas2022MNRAS}.

Figure \ref{fig:selfgravity} shows the gas morphology in simulation L2.0-sg at $t=0.75$~Myr. This is directly comparable to the bottom right plot of Fig. \ref{fig:all_morphology}. Overall, the outflow is remarkably similar, although slightly smaller due to self-gravity effectively increasing the gravitational potential by a factor of $1.1$. This leads to a corresponding decrease in outflow velocity. The gas clumps are slightly smaller and denser, making them more resilient to feedback. As a result, the mass of infalling cold gas is slightly higher, and some clumps fall into the SMBH earlier than in the fiducial simulations. Nevertheless, the simulations evolve qualitatively similarly, leaving our main conclusions unaffected.

Some of the dense self-gravitating gas turns into star particles at approximately constant rates\footnote{We refrain from calling this the star formation rate due to the crudeness of our star formation prescription. The star particles are best thought of as molecular cloud fragments that should form stars, but the fraction of the star particle mass that ends up in stars can range from a few per cent to $\sim 30\%$.} throughout the simulations. By $t = 1$~Myr, the total mass of star particles is $M_*/\left(10^6\,\msun\right) = 1.4, 5.9, 12$, and $17$ in simulations L0-sg, L0.5-sg, L1.0-sg, and L2.0-sg, respectively. The positive correlation between star particle mass and AGN luminosity qualitatively agrees with many other simulations \citep[e.g.][]{Nayakshin2012MNRASb, Gaibler2012MNRAS, Silk2013ApJ, Zubovas2013MNRAS, Zubovas2014MNRASa, Schaye2015MNRAS, Zubovas2017MNRAS, Lauzikas2024AA} and observations \citep[e.g.][]{Cresci2015AA, Maiolino2017Natur, Gallagher2019MNRAS, Shin2019ApJ}.

The results reveal an interesting relationship between SMBH feeding and star formation. In low-luminosity simulations, star particle masses are insignificant compared to infalling cold gas. However, in the two high-luminosity simulations, total star particle mass exceeds that of infalling cold gas. This suggests that star formation in outflow-compressed dense gas can reduce the rate of SMBH feeding and lead to a smaller final SMBH mass. Such a tradeoff between SMBH and stellar mass growth relates to competitive feedback and accretion in galaxy formation \citep{Nayakshin2009MNRAS} and leads to a somewhat counter-intuitive conclusion \citep[see also][]{Silk2013ApJ}: galaxies experiencing extreme, perhaps super-Eddington, AGN episodes at early cosmic times may end up with under-massive SMBHs compared to those that grow at more modest rates. Such a connection would be very important for constraining the models of first SMBH growth at redshifts $z > 6$. It should be checked with dedicated simulations, which we plan to perform in the future.

\subsection{Coexistence of hot and cold flows}

The hot diffuse gas expands much faster than the cold. When such a system is observed, producing a static snapshot, we expect to see ionised gas at larger radii than molecular gas. This is generally the case when we consider large observed outflow samples, i.e. ionised outflows typically have larger extents than neutral or molecular ones (see the compilations in e.g. \citealt{Fiore2017AA}, \citealt{Lutz2020AA}, and \citealt{Fluetsch2021MNRAS} and Fig. 2 in \citealt{Lauzikas2024AA}). The hottest plasma component has recently been probed using both thermal \citep{Hall2019MNRAS} and kinetic Sunyaev-Zeldovich effects \citep{Hadzhiyska2024arXiv}; it was found to extend farther than the dark matter component of galaxies, which is larger than almost all molecular and ionised outflow radii. In individual objects, the data are scarce, but ionised outflows are wider than molecular in PDS 456 \citep{Travascio2024AA}, several local  ultra-luminous infrared galaxies \citep{Fluetsch2021MNRAS}, and $z\sim2$ quasars \citep[although the trend is not universal, with at least one system showing a much larger molecular outflow extent than ionised]{Vayner2021ApJ}.

\citet{Marconcini2025NatAs} recently detected evidence of ionised outflow acceleration beyond a few kiloparsecs in ten local AGNs. This is most likely the result of hot gas outflows escaping the gaseous bulge \citep{ZubovasTartenas2025}. Once this happens, the hot gas pressure decreases significantly, leading to less efficient cold gas clump confinement. This can result in filament expansion, which makes them more susceptible to AGN feedback, so SMBH feeding decreases as well (but see Sect. \ref{sec:magnetic}). Hot gas escape requires that the AGN episode lasts at least $t_{\rm ep,min} \sim 0.5 R_{\rm bulge}/v_{\rm out,hot} \sim 0.5 R_{\rm kpc} v_8^{-1}$~Myr, where $R_{\rm bulge} \equiv 1R_{\rm kpc}$~kpc is the bulge radius and $v_{\rm out,hot} \equiv 1000v_8$~km~s$^{-1}$ is the hot outflow velocity. The factor $0.5$ accounts for fossil outflow expansion after the episode \citep{King2011MNRAS, Zubovas2023MNRAS}. This timescale is only slightly longer than the expected typical AGN episode durations \citep{King2015MNRAS, Schawinski2015MNRAS}. So the changing influence of the hot gas on the cold filaments can be important for regulating SMBH feeding. In future, we plan to enhance our simulations by coupling AGN luminosity to the feeding rate, investigating the self-regulation aspect.

\subsection{Impact of magnetic fields} \label{sec:magnetic}

We neglected the influence of magnetic fields in our simulations. It is known that starburst-driven outflows can lift magnetic fields away from galactic discs \citep{Lopez-Rodriguez2021ApJ}, and the same may be true for AGN-driven outflows and magnetic fields in galactic centres. The hot gas bubbles create loops of magnetic field that can compress the dense filaments in between by magnetic pressure. Additionally, magnetic field enhances the survival of cold gas \citep{Shin2008ApJ, Sparre2020MNRAS}, especially when combined with radiative cooling \citep{Hidalgo-Pineda2024MNRAS}. Thus, magnetic fields can facilitate SMBH feeding by maintaining narrower and denser filaments for longer. As a result, we expect that in a sample of galaxies with a similar velocity dispersion, those with stronger magnetic fields should have more massive SMBHs. It is known that galaxies with stronger magnetic fields have higher star formation rates \citet{Chen2025ApJ} and higher dynamical masses \citet{Tabatabaei2016ApJ}, although correlations with SMBH mass and/or velocity dispersion have not been investigated, to the best of our knowledge. As the number of measurements of galactic magnetic fields grows, this prediction can provide a useful test of our model.

\subsection{Comparison with other work}

Several previous papers investigated the coupling between multi-phase gas and AGN outflows, generally making similar conclusions to ours. We compare our work with them below.

\citet{Zubovas2014MNRASb} showed that dense gas in an outflow receives energy input consistent with momentum driving, while diffuse gas is driven by the whole shocked wind energy. The physical model of feedback used in that paper is equivalent to ours; however, the initial conditions were much simpler, with an initially smooth gas distribution and only an azimuthal density gradient used to create different outflow phases. Furthermore, that paper did not investigate the feeding of the SMBH or gas removal from the nuclear regions.

\citet{Bieri2017MNRAS} investigated the evolution of outflows in clumpy media and found that the momentum loading factor (in their terms, mechanical advantage) decreases as the outflow escapes from the initial cloud and only the dense clumps remain exposed to the AGN feedback. In this case, feedback was mediated by AGN radiation pressure on dusty gas and the authors did not investigate the effect on feeding the SMBH.

More recently, \citet{Ward2024MNRAS} ran simulations of AGN wind driving an outflow through a clumpy disc. The underlying physical model is the same as ours. Again, they found that the dense gas is driven essentially by AGN wind momentum, while the diffuse gas forms an approximately energy-conserving outflow. They did not investigate the integrated effect on gas clusters or the effect on SMBH feeding directly. Their results agree very well with ours, despite them using a different wind velocity ($10^4$~km~s$^{-1}$ instead of $3\times10^4$~km~s$^{-1}$), cooling prescription and numerical scheme (moving mesh instead of SPH).

\section{Summary} \label{sec:sum}

We simulated AGN wind-driven outflows in a turbulent multi-phase medium to analyse the effect of feedback on dense gas and SMBH feeding. Our simulations covered constant AGN luminosities in the range $L_{\rm AGN} = \left(1.3-32\right)\times10^{45}$~erg~s$^{-1}$, corresponding to $0.1-2.5$ times the Eddington luminosity for a black hole on the $M-\sigma$ relation given our galaxy setup, and stochastic variations in initial conditions. We show that dense gas is pushed almost exclusively by the AGN wind momentum, so SMBH feeding is suppressed only when $L_{\rm AGN} \gtrsim 0.7 L_{\rm Edd}$. Since dense gas must be removed far enough to prevent re-accretion after the AGN episode ends, the SMBH mass can grow to or slightly exceed the $M-\sigma$ relation prediction. Additionally, even at very high luminosities, some dense gas filaments lag behind the outflow, although the total mass of dense infalling gas decreases exponentially with AGN luminosity.

Our results align with numerous recent simulations that reveal a complex picture of multi-phase AGN outflows. A single feedback mechanism --- a radiatively driven quasi-relativistic wind, with kinetic power equal to $\sim 0.05$ of the AGN luminosity --- produces outflows spanning wide ranges of radii, gas densities, temperatures, and velocities. Outflow properties can also vary substantially due to differences in host galaxy characteristics, particularly ISM clumpiness and turbulence. We intend to explore these multifaceted AGN--outflow connections in future work, building a consistent framework of AGN feedback applicable to galaxies of different masses, morphologies, gas densities, redshifts, and other relevant parameters.

\begin{acknowledgements} 
MT and KZ are funded by the Research Council Lithuania grant no. S-MIP-24-100. Some simulations were performed on Galax, the computing cluster of the Centre for Physical Sciences and Technology in Vilnius, Lithuania.
\end{acknowledgements}

\bibliographystyle{mnras}
\bibliography{zubovas}

\begin{appendix}

\section{AGN episode and feedback injection} \label{sec:injection}

The AGN affects the gas in two ways: by heating (described in Sect. \ref{sec:sims}) and by producing feedback in the form of a fast wind, which we tracked using a novel grid-based scheme, \texttt{gridWind}. The \texttt{gridWind} method works by propagating the AGN wind effect on a static grid. This allows the wind to be rapidly coupled with nearby SPH particles through a spatial hash-like method and a sorted distance matrix. 

\subsection{Feedback injection with \texttt{gridWind}}

We used a simplified version of \texttt{SREAG} grid \citep{SREAG} with rectangular cells. To construct the grid, we subdivided the sphere's surface into latitudinal strips of a constant $\Delta \theta = \pi / N_{{\rm strip}}$. Each of these strips was then subdivided into a number of rectangular cells, $N_{\Phi}(\theta)$:
\begin{equation}
    N_{\Phi}(\theta) = \left\lfloor \frac{2 \pi \cos({b(\theta)}) }{ \Delta \theta} \right\rceil
    ,\end{equation}
where
\begin{equation}
     b(\theta) = \left\lfloor \frac{\theta}{\Delta \theta} \right\rfloor \Delta \theta + \Delta \theta / 2 - \pi / 2.
\end{equation}
Here $\lfloor x \rfloor$ is the floor function and $\lfloor x \rceil$ denotes rounding to the nearest integer; $\theta$ is the latitude of the midpoint of a given strip. Now the longitudinal size of each rectangular cell in a given strip is $\Delta \phi(\theta) = 2\pi / N_{\Phi}(\theta)$. This results in a deviation of cell area of $\pm1$~\% from the mean except for the two polar strips, where the deviation is about $5$~\%; this is taken into account when distributing feedback instead of modifying the grid itself.  

We can quickly determine the indices ($n_{r}, n_{\theta}, n_{\phi}$) of the nearest grid cell for a given SPH particle at spherical coordinates $r_{\rm p}, \theta_{\rm p}, \phi_{\rm p}$:
\begin{align}
    n_{r} &=\left\lfloor r_{\rm p} / \Delta r\right\rfloor, \\
    n_{\theta} &= \left\lfloor\theta_{\rm p} / \Delta \theta\right\rfloor, \\
    n_{\phi} &= \left\lfloor\phi_{\rm p} / \Delta \phi(\theta)\right\rfloor,
\end{align}
where $\Delta r, \Delta \theta$, and $\Delta \phi$ are the grid step sizes in the three directions. In practice, we derived a more practical set of indices for use in flattened arrays for each shell ($n_{\rm{shell}}$) and the whole spherical volume ($n_{\rm{sphere}}$):
\begin{align}
    n_{\rm{shell}} &= n_{\phi} + \sum_{k=0}^{n_{\theta}} N_{\phi}(\theta),\\
    n_{\rm{sphere}} &= n_{\rm{r}} N_{\rm shell} + n_{\rm{shell}},
\end{align}
where $N_{\rm shell}$ is the number of cells in each shell.  In this work we used $N_{{\rm strip}} = 64$,  which corresponds to $N_{{\rm shell}} = 5216$. 

We used a simple discrete-step approach for wind propagation. We first injected the wind into the zeroth shell in proportion to the total energy produced by the AGN with luminosity $L_{\rm AGN}$ over the SMBH timestep; the amount injected into each cell is weighted by its area. We assumed that the wind travels radially outwards at a constant velocity, $v_{\rm wind}=0.1\rm{c}$ \citep{King2003ApJ, King2010MNRASa}. We equalised the SMBH and wind timesteps by applying an additional constraint, $ \Delta t < C \Delta r / v_{\rm wind}$, where $C=0.4$ is a Courant-type factor. 

Feedback is distributed to SPH particles in proportion to those particles' contribution to the overall density field at the centre of the grid cell. As each SPH particle has spatial extent expressed as the smoothing length $h$, we used a pre-calculated sorted distance matrix to quickly iterate over the nearest neighbouring cells. To prevent unnecessary steps, we checked that the distance from each cell centre is not greater than $h$. We found that, in practice, the radial extent of each particle over different shells can be safely neglected, greatly improving performance. This is safe as long as we only use the ratio between the contribution of the different particles to the overall density field. Each wind variable - in our case, energy $\Delta E = \eta/2 \times E_{\rm cell}$ and momentum $\Delta p = E_{\rm cell}/c$ - is transferred independently. When injecting momentum, the direction of injection is radial if the particle position is inside the cell; otherwise, the direction is parallel to the radial direction of the cell centre.

\begin{figure}
\includegraphics[width=0.49\textwidth, trim={0 10 0 0}, clip]{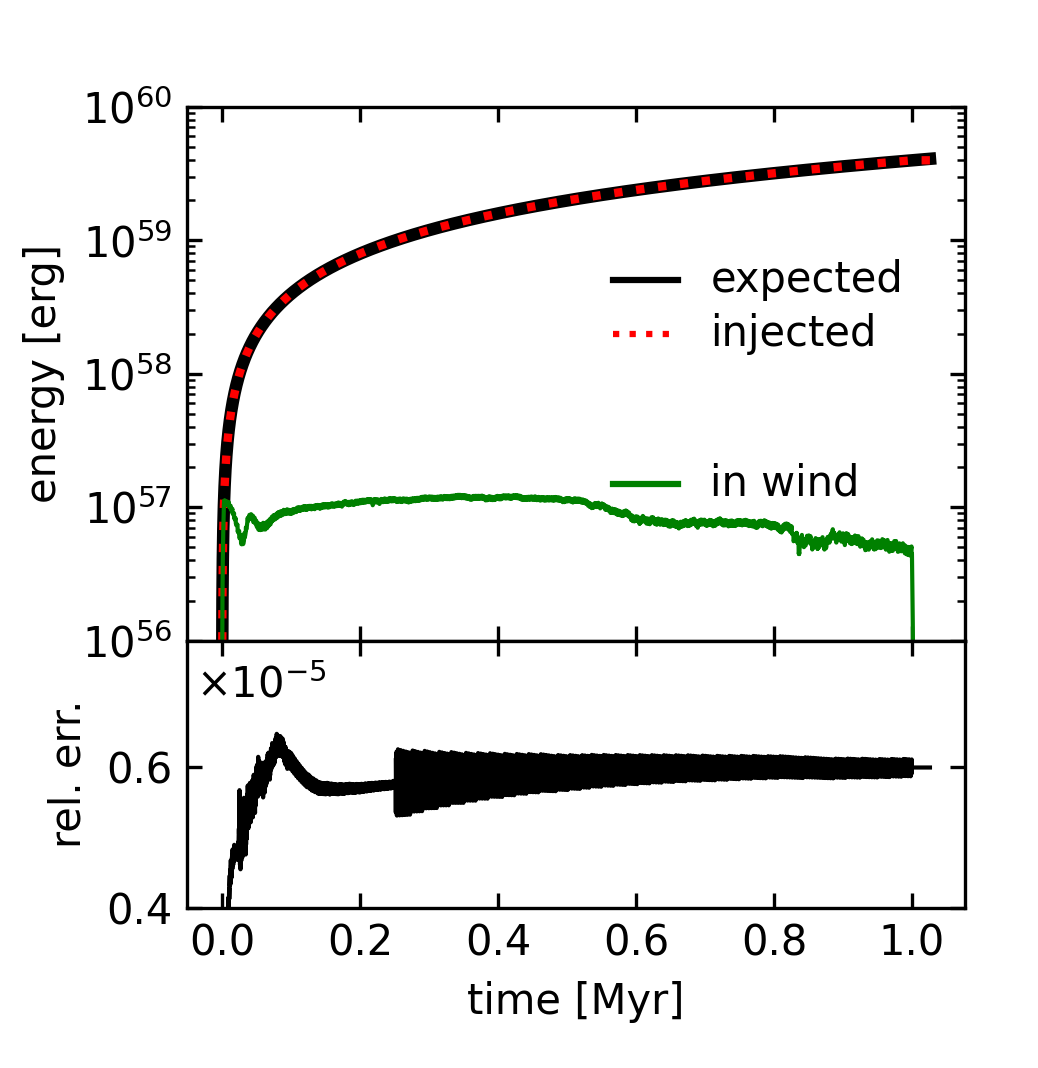}
\caption{Top: Energy injection over time in simulation L1.0. The solid black line shows the expected injection, $E_{\rm tot, exp} = L_{\rm AGN} \times t$. The solid green line shows $E_{\rm w}$, the amount of energy contained within the wind that is yet to be injected. The dotted red line shows $E_{\rm inj}$, the actual amount of energy injected into the particles divided by $\eta/2$. Bottom: Relative error, $\Delta_{\rm inj} = \left| E_{\rm tot, exp} - \left( E_{\rm inj} + E_{\rm w} \right) \right|/E_{\rm tot, exp}$. }
\label{fig:windError}
\end{figure}

We tracked the energy injection over the entire run at each injection step in a log file; an example of the output from one L1.0 simulation is shown in Fig. \ref{fig:windError}. Comparing the black and dotted red lines, we see an excellent agreement, with relative error approximately constant at $6\times10^{-6}$. Momentum injection is equally precise because both quantities are proportional to the energy contained within a grid cell, which is the quantity tracked in the plot.

\subsection{Momentum-only wind test} \label{sec:momtest}
\begin{figure}
\includegraphics[width=0.49\textwidth, trim={0 0 0 0}, clip]{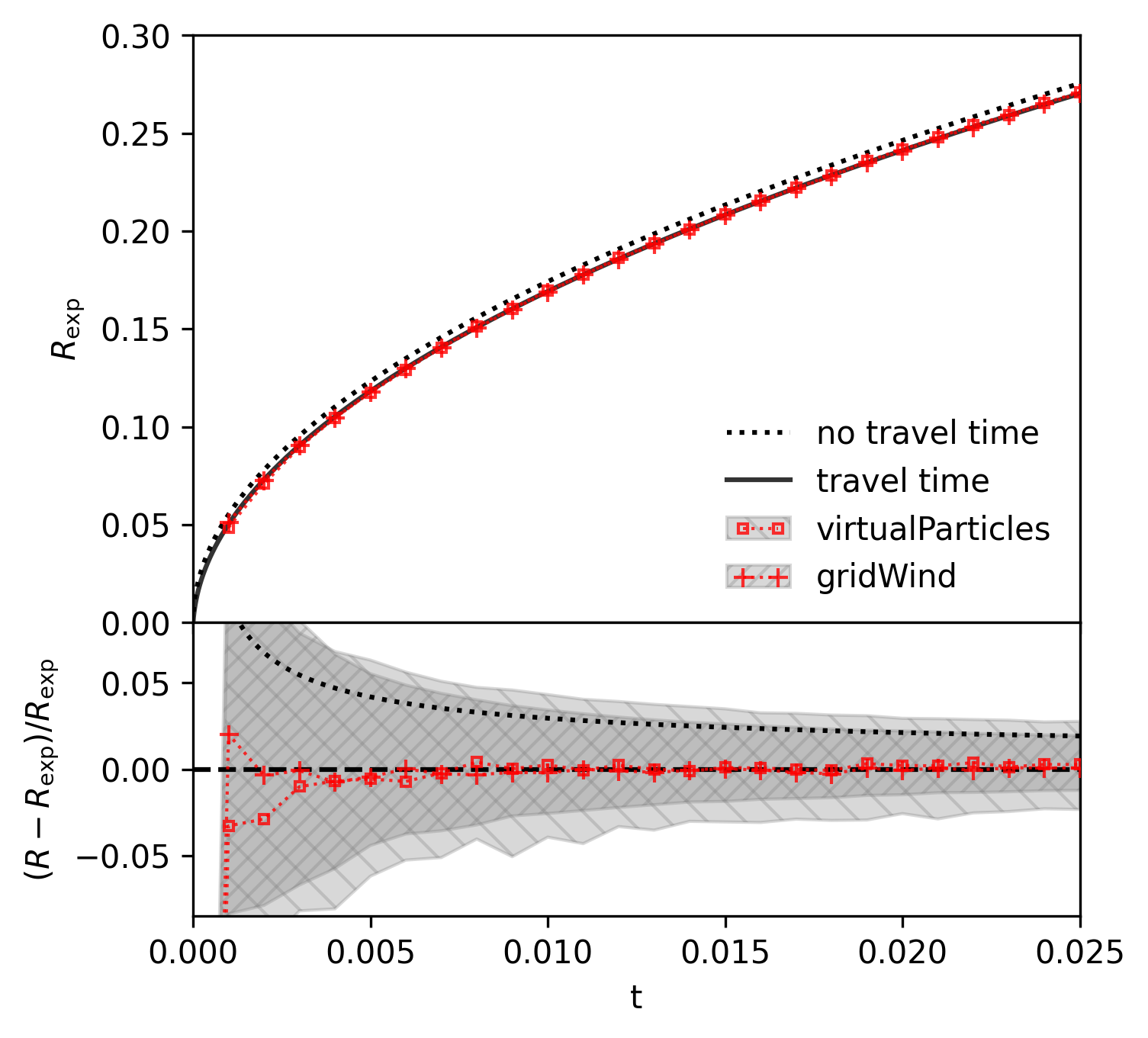}
\caption{Top: Evolution of the radius of a sphere expanding into uniform isothermal gas due to momentum injection. Bottom: Deviation from the analytical solution. Shaded areas correspond to the width at half maximum from the density peak.}
\label{fig:momWindTest}
\end{figure}

\begin{figure}
\includegraphics[width=0.49\textwidth, trim={0 0 0 0}, clip]{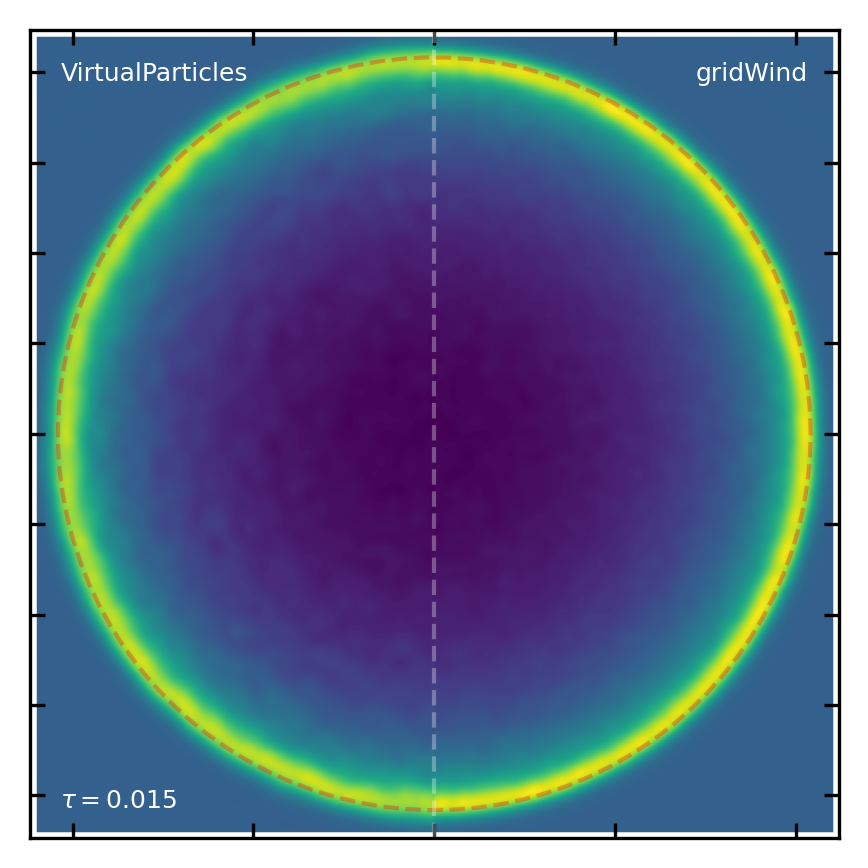}
\caption{Density maps of the resultant spheres. The red dashed line shows the expected radius.}
\label{fig:momWindCompare}
\end{figure}

We illustrate the viability of our approach by performing the momentum-driven wind test as outlined in \cite{Nayakshin2009MNRAS}. In this setup, a luminous source is placed at the centre of a unit periodic box containing isothermal gas of constant density. Feedback is injected spherically into the surrounding medium in the form of outward momentum resulting in an expanding central cavity. Equating the rate of change of momentum of the expanding cavity to the momentum input from the luminous source gives
\begin{equation}
\frac{d}{dt} \left[ \left( \frac{4}{3} \pi R^3 \rho_0 \right) \dot{R} \right] = \frac{L}{c},
\label{eq:m}
\end{equation}
which is solved for the radius of the resultant cavity, $R(t)$:
\begin{equation}
R(t)=\left( \frac{3L}{2 \pi c \rho_0} \right)^{1 / 4} t^{1/2},
\label{eq:revs}
\end{equation}
where $L$ is the luminosity of the central source, $c$ is the speed of light and $\rho_0$ is the initial density of the surrounding medium. A more complex estimate takes into account the finite velocity of the wind $v$ and the restoring force due to external pressure: 
\begin{equation}
\left[\left(\frac{4}{3} \pi R^3 \rho_0\right) \dot{R}\right]=\frac{L}{c}\left(t-\frac{R}{v}\right)-4 \pi R^2 \rho c_{\rm s}^2 t, 
\label{eq:revc}
\end{equation}
where $c_s$ is the speed of sound. This expression is solved for $R(t)$ numerically.

We performed two tests with \texttt{Gadget}, using $N_{\rm sph}=1\times10^6$ particles: one with virtual particles \citep{Nayakshin2009MNRAS} and one with \texttt{gridWind}. For \texttt{gridWind} we used $N_{{\rm strip}}=64$, which corresponds to $N_{{\rm shell}}=5216$ (the same as the main set of simulations) and the number of radial shells $N_{\rm r}=200$, with $\Delta r = 0.0025,$ which is of the same order as the minimum smoothing length $h=0.001$. For virtual particles we used reasonably good resolution with $p_{\rm \gamma}=0.5 m_{\rm sph} c_{\rm s}$. The resulting expansion of the central cavity is shown in the top panel of Fig. \ref{fig:momWindTest}. After the initial few steps, both approaches converge on the expected values given by Eq. (\ref{eq:revc}). In the bottom panel the agreement is highlighted by showing the relative difference from the analytical solution tracing both the density peak and its width. Our method produces a slightly finer peak, although this is more sensitive to the overall SPH simulation resolution. A density map of the spheres in both simulations (Fig. \ref{fig:momWindCompare}) confirms that \texttt{gridWind} results in a slightly smoother density distribution with a more defined peak at the correct radius.

\end{appendix}

\end{document}